\documentclass[11pt]{article}
\usepackage{graphicx}
\usepackage{amsmath}
\usepackage{amssymb}
\usepackage{amsxtra}
\usepackage{color}

\providecommand{\gtrless}{
  \mathrel{
  \smash{
  \vcenter{
    \offinterlineskip 
    \ialign{
       \hfil##\hfil\cr 
       $>$\cr 
       \noalign{\kern-.3ex}
       $<$\cr 
    }
  }
  }
  \vphantom{>}
  }
}

\begin{document}

\begin{flushright} 
OIQP-15-9
\end{flushright} 

\begin{Large}
\vspace{1cm}
\begin{center}
Bosons being their own antiparticles in Dirac formulation
\end{center}
\end{Large}

\begin{center}
{\large Holger Bech Nielsen$^{a)}$

and Masao Ninomiya$^{b)}$}

\vspace{0.4cm}

a)
Niels Bohr Institute, Copenhagen University,
\\Blegdamsvej 17, 2100 Copenhagen $\phi$, Denmark 
\\E-mail: 
hbech@nbi.dk, 
hbechnbi@gmail.com\\

b)
Okayama Institute for Quantum Physics,\\
Kyoyama 1-9-1 Kita-ku, Okayama-city 
700-0015, Japan\\
E-mail: msninomiya@gmail.com

\end{center}

\begin{abstract}
Using our earlier formalism of extending the idea 
of the Dirac sea (negative energy states) to also for 
Bosons, we construct a formalism for Bosons which 
are their own antiparticles.  Since antiparticles 
in formalisms with a Dirac sea are at first formulated 
as holes, they are a priori formally a bit different 
from the particles themselves.  To set up a formalism/a 
theory for Majorana fermions and for Bosons in which particles are 
their own antiparticles is thus at first non-trivial.  
We here develop this not totally trivial formalism 
for what one could call extending the name for the fermions, 
``Majorana-bosons''.  Because in our earlier work had what we 
called ``different sectors'' we got there some formal 
extensions of the theory which did not even have positive 
definite metric.  Although such unphysical sectors may 
a priori be of no physical interest one could hope that 
they could be helpful for some pedagogical deeper 
understanding so that also the formalism for particles 
in such unphysical sectors of their own antiparticles 
would be of some ``academic'' interest.

\end{abstract}

\begin{flushleft}
PACS NO: 
11.10.-z (Field Theory),
11.25.-w (String and Branes),
14.80.-j (other particles (including hypothesical particles)),
03.70.+K (Theory of Quantized Fields).

Keywords: Dirac sea, Majorana fermion, 
Dirac sea for Bosons, Quantum field theory,
String field theory
\end{flushleft}

\section{Introduction}

Majorana \cite{Majorana} put forward the 
idea of fermions 
having the property of being their own 
antiparticles and such
fermions are now usually called Majorana 
fermions or Majorana
particles.

The Majorana field $\psi_M$ is defined in general as real or hermitean
$\psi^{\dagger}_{M}=\psi_M$.

Among the known bosons we have more 
commonly bosons, which are 
their own antiparticles, and which we 
could be tempted to
 call analogously ``Majorana bosons''
(a more usual name is ``real neutral particles'' \cite{realneutral}),
 such as the photon, 
 $Z^0$, $\pi^0$, $\ldots$ particles.  Now 
the present authors extended 
 \cite{Diracsea, wHabara} the Dirac sea idea 
\cite{Dirac} of having negative energy 
 electron single particle states in the second quantized theory 
 being already filled in vacuum, also to Bosons.  This extension 
 of the Dirac sea idea to Bosons has a couple of new features:
\begin{description}
\item[1)] 

 We had to introduce the concept of having a negative number of 
 bosons in a single particle state.  We described that by 
 considering the analogy of a single particle state in which 
 a variable number of bosons can be present to a harmonic 
 oscillator, and then extend their wave functions from normalizable 
 to only be analytical.  The harmonic oscillator with wave functions 
 allowed to be non-normalizable and only required to be analytical 
 has indeed a spectrum of energies $E_n=
(n+\frac{1}{2})\omega$ 
 where now $n$ can be all integers $n=\ldots, -3,-2,-1,0,1,2,\ldots$.
So it corresponds to that there can be a negative number of bosons 
in a single particle state.
\item[2)]
It turns out though that these states - of say the ``analytical 
wave function harmonic oscillator'' corresponding to negative 
numbers of bosons have alternating norm square:
For $n\ge 1$ we have as usual $\langle{n|n}\rangle=1$ for $n\ge 0$ 
(by normalization) but for $n\le -1$ we have instead 
 $\langle{n|n}\rangle=c\cdot(-1)^n$ for $n$ negative.
($c$ is just a constant we put say $c=+1$.)  
This variation of norm square is needed to uphold the usual rules 
for the creation $a^+$ and annihilation $a$ operators

\begin{align}
a^+|n\rangle&=\sqrt{1+n}|n+1\rangle \nonumber \\
a|n\rangle&=\sqrt{n}|n-1\rangle \label{car}
\end{align}
to be valid also for negative $n$.
\item[3)]
With the relations (\ref{car}) it is easily seen that there is a 
``barrier'' between $n=-1$ and $n=0$ in the sense that the 
creation and annihilation operators $a^+$, and $a$ cannot 
bring you across from the space spanned by the $n=0,1,2,\ldots$ 
states to the one spanned by the $n=-1,-2,\ldots$ one 
or opposite.  It is indeed best to consider the usual space spanned 
by the $|n\rangle's$ with $n=0,1,2,\ldots$ as one separate ``sector'' 
the ``positive sector'' and the one spanned by the $|n\rangle$ states 
with $n=-1,-2,\ldots$ as another ``sector'' called the 
``negative sector''.  Since in the harmonic oscillator with 
the wave functions only required to be analytical but not 
normalizable the states in the ``positive sector'' are not truly 
orthogonal to those in the ``negative sector'' but rather 
have divergent or ill-defined inner products with each other, 
it is best not even to (allow) consider inner products like say

\begin{align}
\langle 0|-1\rangle&=\mbox{ill defined} \nonumber \\
\langle n|p\rangle&=\mbox{ill defined} 
\end{align}

when $n\le -1$ and $p\ge0$ or opposite.  
\\Basically we shall 
consider only \underline{one sector} at a time.
\item[4)]
The use of our 
formalism
with negative number of particles 
to connect to the usual and physically correct description of 
bosons with some charge or at least an (at first) conserved 
particle number comes by constructing a ``Dirac sea for bosons''.  
That is to say one first notes that e.g. the free Klein-Gordon equation
\[\Box \phi =0\]
has both positive and negative energy solutions, 
and that the inner product
\begin{equation}
\langle\varphi_1|\varphi_2\rangle=\int \varphi^\ast_1\overleftrightarrow{\partial}_0\varphi_2d^3\vec{X}
\end{equation}
gives \underline{negative norm square for negative 
energy eigenstates} and positive norm square for positive 
energy eigenstates.\\
Then the physical or true world is achieved by using for
the negative energy single particle states the ``negative sector''
(see point 3) above) while one for the positive energy single particle 
states use the ``positive sector''.  That is to say that in the
physical world there is (already) a \underline{negative number of
bosons} in the negative energy single particle states.  In the 
vacuum, for example, there is just $-1$ boson in each negative
energy single particle state.
\\This is analogous  to that for fermions 
there is in the Dirac sea just $+1$ fermion in each negative energy
single particle state.  For bosons --  where we have $-1$ instead $+1$ particle--  
we just rather emptied Dirac sea by one boson in each single particle 
negative energy state being removed from 
a thought upon  situation with  
with $0$ particles everywhere.
(Really it is not so nice to think on this removal because the
``removal'' cross the barrier from the ``positive sector'' to the
``negative sector'' and strictly speaking we should only look at
one sector at a time (as mentioned in 3).)
\item[5)]
It is rather remarkable that the case with
the ``emptied out Dirac sea'' described in 4)
-- when we keep to positive sector for 
positive energy and negative 
sector for negative energy-- we obtain a positive definite Fock space.
This Fock space also has only positive energy of its excitations as
 possibilities.  Indeed we hereby obtained exactly a Fock space for
a theory with bosons, that are different from their antiparticles.

\end{description}

In the present paper we like to study how to present a theory
for bosons which are their own antiparticles, Majorana bosons so 
to speak, in this formalism with the ``emptied'' Dirac sea.\\

Since in the Dirac sea formalisms -- both for fermions and for 
bosons -- an antiparticle is the removal of a particle from the 
Dirac sea, an antiparticle a priori is something quite different from
a particle with say positive energy.  Therefore to make a theory / a
formalism for a theory with particle being identified with its
antiparticles -- as for Majorana fermions or for the 
photon, $Z^0$, $\pi^0$ -- is in our or Dirac's Dirac sea formalisms
a priori not trivial. Therefore this article.  Of course it is at the
end pretty trivial, but we think it has value for our understanding
to develop the formalism of going from the Dirac sea type picture
to the theories with particles being their own antiparticles
(``Majorana theories'').

One point that makes such a study 
more interesting is that we do \underline{not} have to only
consider the physical model in the boson case with using positive
sector for positive single particle energy states and negative sector
for negative energy single particle states.  Rather we could -- as a
play-- consider the sectors being chosen in a non-physical way.
For example we could avoid ``emptying'' the Dirac sea in the boson-case
and use the positive sector for both negative and positive 
energy single boson eigenstates.  In this case the Fock space would not
have positive norm square.  Rather the states with an odd number of negative
energy bosons would have negative norm square, and  of course allowing a positive
number of negative energy bosons leads to their being no bottom in the
Hamiltonian for such a Fock space.

The main point of the present
article is to set up a formalism for particles that are their own antiparticles
(call them ``Majorana'') on the basis of a formalism for somehow charged particles 
further formulated with the Dirac sea.  That is to say we consider 
as our main subject how to restrict the theory with the Dirac sea -- and at
first essentially charged particle -- to a theory in which the particles
and antiparticles move in the same way and are identified with each other.

For example to describe a one-particle state of a ``Majorana'' particle one
would naturally think that one should use a state related to either 
the particle or the antiparticle for instance being a superposition
of a particle and antiparticle state.

So the states of the
Fock space $H_{Maj}$ for describing the particles which are their own
antipatricles shall be below identified with some corresponding states
in the theory with Dirac sea.  However, there are more degrees of 
freedom in a theory with charged particles (as the Dirac sea one) than in
a corresponding theory for particles which are their own antiparticle.  
Thus the states in the with Dirac sea Fock space cannot all be transfered
to the Fock space from ``Majorana'' particles.  So only a certain subspace of
the Fockspace for the with Dirac sea theory can be identified with
states of some number of Majorana particles.

To develop our formalism for
this transition from the Dirac sea theory to the one for Majorana particles,
we therefore need a specification of which subspace is the one to be used
to describe the ``Majorana particles''.  Below we shall argue for that this
subspace $H_{Maj}$ becomes  
\begin{equation}
H_{Maj}=\left\{|\hspace{0.2cm}\rangle|\left(a(\vec{p},E>0)+a^\dagger(-\vec{p},E<0)\right)|\hspace{0.2cm}\rangle=0, \mbox{for all}\ \vec{p} 
 \right\}
\end{equation}
\\where we used the notation of $a(\vec{p},E<0)$
for the annihilation operator for a particle with momentum $\vec{p}$ and
energy $E$ corresponding to that 
being positive 
i.e. $E>0\Rightarrow E=\sqrt{m^2+\vec{p^2}}$.
Correspondingly the annihilation operator  $a(\vec{p},E<0)$
annihilates a particle with energy $ E=-\sqrt{m^2+\vec{p^2}}$.  The corresponding
creation operators just have the dagger $\dagger$\ attached to the annihilation
operator, as usual.  We define
\begin{equation}
r(\vec{p})=\frac{1}{\sqrt{2}}\left(a(\vec{p},E>0)+a^{\dagger}(-\vec{p},E<0)\right) \label{r(vec{p})=}
\end{equation} 

This then shall mean, that we should identify a basis, the basis elements of which have
a certain number of the ``Majorana bosons'', say, with some momenta --  in a 
physical world we only expect conventional particles with positive energy -- for
the subspace  $H_{Maj}$ contained in the full space with Dirac sea.

We thus have to construct below creation $b^\dagger(\vec{p})$ and
annihilation $b(\vec{p})$ operators
for the particles which are their own antiparticles (``Majoranas'').  These
$b^\dagger(\vec{p})$ and $b(\vec{p})$ should now in our work be presented by formulas 
giving them in terms of the creation and annihilation operators for the theory
with Dirac sea (and thus acting on the Fock space $H$ of this ``full''
theory).  In fact we shall argue for (below)
\begin{equation}
b^\dagger_n=\begin{cases}
\frac{1}{\sqrt{2}}(a^{\dagger}\left(a^{\dagger}\vec{p},E>0)-a(-\vec{p},E<0)\right)\ (on\ pos.\ sec\ for\ pos.\ E,\ neg\ sec\ for\ neg\ E)\\
\frac{1}{\sqrt{2}}(a^{\dagger}\left(a^{\dagger}\vec{p},E>0)+a(-\vec{p},E<0)\right)\ for\ both\ sectors\\
\cdots 
\end{cases}
\end{equation}
\\and then of course it has to be so that these $b^\dagger(\vec{p})$  and $b(\vec{p})$ 
do not bring a Hilbert vector out of the subspace $H_{Maj}$ but let it
stay there once it is there.  It would be the easiest to realize such a 
keeping inside $H_{Maj}$ by action with 
$b^\dagger(\vec{p})$ -- and we shall
have it that way -- if we arrange the commutation rules
\begin{align}
\left[r(\vec{p}),b^\dagger (\vec{p'})\right]&=0\nonumber
\\\left[r(\vec{p}),b(\vec{p'})\right]&=0
\end{align} 

(Here the commutation for $\vec{p}\neq \vec{p'}$ is trivial because it then concerns
different d.o.f. but the 
$\left[r(\vec{p}),b^\dagger (\vec{p'})\right]=0$ and
$\left[r(\vec{p}),b(\vec{p'})\right]=0$ 
are the nontrivial relations to be arranged (below))

Indeed we shall find below
\begin{align}
b^{\dagger}(\vec{p})&=\frac{1}{\sqrt{2}}  
\left(a^{\dagger}(\vec{p},E>0)+a (-\vec{p},E<0)\right)\ \ (defined\ on\ both\ pos.)\nonumber
\\b(\vec{p})&=\frac{1}{\sqrt{2}}\left(a(\vec{p},E>0)+a^{\dagger}(-\vec{p},E<0\right)
\end{align}

It is then that we shall arrange that if we extrapolate to define also the 
$b^{(\dagger)}(\vec{p},E<0)$
and not only for positive energy $b^{(\dagger)}(\vec{p},E>0)=b^{(\dagger)}(\vec{p})$
we should obtain the formula usual in conventional description of 
Majorana particle theories

\begin{align}
b^\dagger(\vec{p})&=b^\dagger(\vec{p},E>0)=b(-\vec{p},E<0)\nonumber
\\b(\vec{p})&=b(\vec{p},E>0)=b^\dagger (-\vec{p},E<0)
\end{align}

For fermions we simply do construct these $r(\vec{p})$ and $b(\vec{p})$ rather trivially and
it must be known in some notation to everybody.  For bosons, however almost 
nobody but ourselves work with Dirac sea at all, and therefore it must
be a bit more new to get particles which are their own antiparticles into such
a scheme.  For bosons also we have already alluded to the phenomenon of 
different ``sectors'' (see 3) above) being called for due to our need for negative
numbers of particles.  We therefore in the present article as something also new
have to see what becomes of the theory with 
bosons being their own
antiparticles when we go to the unphysical sector-combinations. (The physical combination 
of sectors means as described in point 4) above, but if we e.g. have the positive sector
both for negative and positive energy single particle states, this is a unphysical sector-combination.)
This is a priori only a discussion though of academic interest, since the truly physical 
world corresponds to the physical combination described in point 4) with
the Dirac sea ``emptied out''.

However, in our attempts to describe string field theory
in a novel way we raised to a problem that seemed formally to have solution 
using such on unphysical sector-combination.\\

In the following section 2 we just, as a little warm up, discuss the 
introduction in the Majorana fermion theory on a subspace of the Fock space of
a fermion theory in Dirac sea formulation.

In section 3 we then review with more formalism our ``Dirac sea for bosons'' theory.

Then in section 4 we introduce the formalism $r(\vec{p}),b^\dagger(\vec{p})$
and $b(\vec{p})$ relevant for the Majorana rather theory or for particles 
which are their own antiparticles.  The operators $r(\vec{p})$ defined in (\ref{r(vec{p})=}) are the operators defined 
to be used for singling out the Majorana subspace, and $b^\dagger (\vec{p})$ and $b(\vec{p})$
are the creation and annihilation operators for ``Majorana-bosons''.

In section 5 we go to the unphysical sector combinations to study the presumably only
of acdemic interest problems there.

In section 6 we bring conclusion and outlook.


\section{Warming up by Fermion}

\subsection{Fermion Warm Up Introduction}

As the warming up consider that we have a fermion theory
at first described by making naively (as if 
nonrelativistically, but we consider relativity) creation
$a^\dagger (n,\vec{p},E>0)$ and $a^\dagger (n,\vec{p},E<0)$ for
respectively positive and negative energy $E$ of the single 
particle state.  Also we consider the corresponding 
annihilation operators $a(\sigma ,\vec{p},E>0)$ and $a(\sigma ,\vec{p},E<0)$
\\The physically relevant second quantized system takes its outset
in the physical vacuum in which all the negative energy 
$E<0$ single particle states are filled while the positive
energy ones are empty.
\begin{equation}
\mid\ vac\ phys\rangle=\prod_{\sigma,\vec{p}} a^{\dagger}(\sigma,\vec{p},E<0)\mid0\ totally\ empty\rangle
\end{equation}

Of course in modern practice you may ignore the
Dirac sea and just start from the physical vacuum
$\mid vac\ phys\rangle$ and operate on that with creation and
annihilation operators.  If you want to 
say create on antiparticle with momentum $\vec{p}$
(and of course physically wanted positive energy)
you operate on $\mid vac\ phys\rangle$ with 
\begin{equation}
a^\dagger_{anti}(\sigma,\vec{p},E>0)=
a(\sigma^1 ,-\vec{p},E<0)
\end{equation}
i.e. the antiparticle creation operator 
$a^\dagger_{anti}(\sigma,\vec{p}, E>0)$ is equal to
the annihilation operator $a(\sigma^1 ,-\vec{p},E<0)$
with the ``opposite'' quantum numbers.

\subsection{Constructing Majorana}

Now the main interest of the present article is how
to construct a theory of particles being
their own antiparticle (``Majorana'') from
the theory with essentially charged particles
-- carrying at least a particle-number
``charge''-- by appropriate projection out of a
sub-Fock space and by constructing creation and
annihilation operators for the Majorana --in this
section-- fermions.

Let us remark that this problem is so simple, that we
can do it for momentum 
value, and if we
like to simplify this way we could decide to consider 
only one single value of the momentum $\vec{p}$ and spin.
Then there would be only two creation and two annihilation
operators to think about
\begin{align}
a^\dagger(E>0)&=a^\dagger(\sigma ,\vec{p},E>0)\nonumber
\\a^\dagger(E<0)&=a^\dagger(\sigma ,\vec{p},E<0)
\end{align}
\\and thus the whole Fock space, we should play with
would only have $2\cdot2 =4$ states, defined by having filled
or empty the two single particle states 
being the only ones considered
in this simplifying description just 
denoted by ``$E>0$'' and ``$E<0$''.

In fact the construction of a full Majorana-formalism will
namely be obtained by making the construction of the
Majorana Fock (or Hilbert) space for each momentum $\vec{p}$
and spin  and then take the Cartesian product of all the
obtained Majorana-Fock spaces, 
a couple 
for each spin and momentum combination.

The four basis states in the Fock space after throwing away all
but one momentum and one spin-state  are:
\begin{align}
\mid 1\ antiparticle\rangle&=\mid vac\ totally \ empty\rangle\nonumber
\\\mid vac\ phys\rangle&=a^\dagger (E<0)\mid vac\ totally\ empty\rangle\nonumber
\\\mid 1\ fermion\ in\ phys\rangle&=a^\dagger(E>0)a^\dagger(E<0)
\mid vac\ totally\ empty\rangle\nonumber
\\\mid both\ particle\ and\ antip.\rangle&=a^\dagger(E>0)\mid vac\ totally \ empty\rangle
\end{align}

Considering the situation from the point of view of the
physical vacuum

\begin{equation}
\mid vac\ phys\rangle=a^\dagger (E<0)\mid vac\ totally\ empty\rangle
\end{equation}
\\creating a Majorana particle should at least either 
a particle or an antiparticle or some superposition
of the two (but not both).

So the one Majorana particle state shoule be a 
superpositon of
\begin{equation}
\mid 1\ fermion\ in\ phys\rangle=a^\dagger(E>0)a^\dagger(E<0)
\mid vac\ totally\ empty\rangle
\end{equation}
\\and
\begin{equation}
\mid 1\ antiferm\ in\ phys\rangle=\mid vac\ totally\ empty\rangle
\end{equation}

The most symmetric state would natutally be to take with
coefficients $\frac{1}{\sqrt{2}}$ these two states
with equal amplitude:
\begin{equation}
\mid 1\ Majorana\rangle=\frac{1}{\sqrt{2}}\left(a^\dagger(E>0)a^\dagger(E<0)
+1\right)\mid vac\ totally\ empty\rangle
\end{equation}

We should then construct a creation operators
$b^\dagger(\sigma,\vec{p})$ or just $b^\dagger$
so that
\begin{equation}
b^\dagger\mid vac\ phys\rangle=\mid 1\ Majorana\rangle
\end{equation}
Indeed we see that
\begin{equation}
b^\dagger=\frac{1}{\sqrt{2}}\left(a^\dagger(E>0)+a(E<0)\right)\label{cref}
\end{equation}
\\will do the job.

If we use $b^\dagger$ and  
\begin{equation}
b=\frac{1}{\sqrt{2}}\left(a(E>0)+a^\dagger(E<0)\right)\label{anf}
\end{equation}
\\it turns out that states needed are the -- superpositoins of --
\begin{equation}
\mid vac\ phys\rangle=a^\dagger(E<0)\mid vac\ totally\ empty\rangle
\end{equation}
\\and
\begin{equation}
b^\dagger\mid vac\ phys\rangle=\mid 1\ Majorana\rangle
\end{equation}

This subspace which in our simplyfication
of ignoring all but one momentum and spin
state actually represents the whole space
$H_{Maj}$ used to describe the Majorana theory
has only \underline{2\ dimensions} contrary
to the full Hilbert space $H$ which in our
only one momentum and spin consideration has
\underline{4\ dimensions}.

So it is a genuine subspace and we shall
look for an operator $r=r(h,\vec{p})$ which
gives zero  when acting on $H_{Maj}$ but not when
it acts on the rest of $H$.

It is easily seen that
\begin{equation}
r=\frac{1}{\sqrt{2}}\left(a(E>0)-a^\dagger(E<0)\right)
\end{equation}
\\will do the job.  Thus we can claim that
\begin{equation}
H_{Maj}=\{\mid\rangle \mid r\mid \rangle =0\}
\end{equation}

Written for the full theory with all the momenta and
spins we rather have
\begin{equation}
H_{Maj}=\{\mid\hspace{0.2cm}\rangle \in H\mid\forall_{h} \vec{p}[r(\vec{p},h)\mid\rangle=0]\}
\end{equation}
\\where
\begin{equation}
r(\vec{p},h)=\frac{1}{\sqrt{2}}\left(a(\vec{p},h,E>0)-a^\dagger(-\vec{p},h,E<0)\right)
\end{equation}
\\and $a(\vec{p},h,E>0)$ is the annihilation operator for a 
fermion with momentum $\vec{p}$ and eigenstate $h$ of the
normalized helicity
\begin{equation}
h\sim \vec{\Sigma}\cdot \vec{p}/\mid\vec{p}\mid
\end{equation}
where $\vec{\Sigma}$ is the spin angular momentum, and
the energy $E=+\sqrt{\vec{p^2}+m^2}$.

The fully described creation opperator for a
Majorana \underline{particle} fermion with
momentum $\vec{p}$ and helicity $h$,
\begin{equation}
b^{\dagger}(\vec{p},h)=\frac{1}{\sqrt{2}}\left(a^{\dagger}(\vec{p},h,E>0)+a(-\vec{p},h,E<0)\right)
\end{equation}
\\and the corresponding annihilation operator
\begin{equation}
b(\vec{p},h)=\frac{1}{\sqrt{2}}\left(a(\vec{p},h,E>0)+a^\dagger(-\vec{p},h,E<0)\right)
\end{equation}

One easily checks that the operation with these operators
$b(\vec{p},h)$ and $b^+(\vec{p},h)$ map $H_{Maj}$ on $H_{Maj}$ 
because
\begin{align}
\left\{r(\vec{p'},h'),b(\vec{p},h)\right\}_+&=0\nonumber
\\\left\{r(\vec{p'},h'),b^{\dagger}(\vec{p},h)\right\}_+&=0
\end{align}


\section{Review of Dirac Sea for Bosons}
\label{Sec 3}


Considering any relativistically invariant
dispersion relation for a single particle
it is, by analyticity or better by having a
finite order differential equation,
impossible to avoid that there will be
both negative and positive energy (eigen)
solutions.  This is true no matter whether you think of
integer or half integer spin or on bosons or
fermions(the latter of course cannot matter
at all for a single particle theory).  In fact this
unavoidability of also negative energy single
particle states is what is behind the unavoidable
CPT-theorem.

There is for each type of equation a corresponding 
inner product for single particle states, so that
for instance the Klein-Gordon equation and the Dirac equation
have respectively 
\begin{equation}
\langle\varphi_1\mid\varphi_2\rangle=\int \varphi^{\ast}_1\frac{\overleftrightarrow{\partial}}{{\partial}t}\varphi_{2}d^3\vec{X}
\ (Klein\ Gordon)\label{inb}
\end{equation}
\\and
\begin{equation}
\langle\psi _1\mid\psi_2\rangle=\int\psi^{\dagger}_1\psi_{2}d^3\vec{X}=\int\vec{\psi}_{1}\gamma ^{0}\psi_{2}d^{3}\vec{X}
\ (for\ Dirac\ equation)
\end{equation}
( see e.g. \cite{inner})
 

At least in these examples --but it works more generally--
the inner product of a single particle state with itself, 
the norm square, gets negative for integer spin and remains
positive for the half integer spin particles, when going to the negative energy 
states. 

For integer spin particles (according to spin statistics theorem
taken to be bosons) as for example a scalar we thus have
\underline{negative norm square} for the \underline{negative
energy} single particle states.  This means that for all the
states for which we want to make an analogy to the filling of
the Dirac sea, we have to have in mind, that we have this negative
norm square.

That is to say, that thinking of second quantizing the norm
square of a multiple particle state in the Fock space would 
a priori alternate depending on whether the number of 
particles (bosons) with negative energy is even or odd.

Physically we do not want such a Fock space, 
which 
has non-positive-definite norm --since for the
purpose of getting positive probabilities we need a 
positive definite inner product --.

The resolution to this norm square problem in 
our ``Dirac sea for bosons'' --model is to compensate
the negative norm square by another negative
norm square which appears, when one puts into a single
particle state a negative number of bosons.

This is then the major idea of our 
`Dirac sea for
bosons''-work, that we formally --realy of course our
whole model in this work is a formal game -- assume that
it is possible to have a negative number of particles
(bosons) in a single particle state.  That is to say we 
extend the usual idea of the Fock space so as to not as
usual have its basic vectors described by putting various
non-negative numbers of bosons into each single particle state, but allow also to
have a negative number of bosons.

Rather we allow also as Fock-space basis vector states
corresponding to that there could be negative integer
numbers of bosons.  So altogether we can have any integer 
number of bosons in each of the single particle states
(whether it has positive or negative energy at first
does not matter, you can put any integer number of bosons
in it anyway).

In our ``Dirac-sea for Bosons'' --paper
\cite{Diracsea} we present the
development to include negative numbers of 
particles via the analogy with an harmonic oscillator.
It is well-known that a single particle state with
a non-negative number of bosons in it is in perfect
correspondance with a usual harmonic 
oscillator\cite{Feynman} 
in which the number of
excitations can be any positive number or zero.
If one extend the harmonic oscillator to have in the
full complex plan extending the position
variable $q$ (say)and the wave function
$\psi(q)$  to be 
formally analytical
wave function only, but give up requiring normalizability,
it turns out that the number of excitations $n$ extends to 
$n\in Z$, i.e. to $n$ being any 
integer.  This analogy to extend harmonic oscillator can be used
to suggest how to build up $a$ formalism withe creation $a^\dagger$
and annihilation operators $a$ and an inner product for a
single particle states in which one can 
have any
integer number of bosons.

It is not necessary to use extended harmonic oscillator.  In fact 
one could instead just write down the usual relations for creation
and annihilation operators first for a single particle state say
\begin{equation}
a^\dagger(\vec{p},E>0)\mid k(\vec{p},E>0)\rangle=\sqrt{k(\vec{p},E>0)+1}\mid k(\vec{p},E>0)+1\rangle \label{ca34}
\end{equation}
\\and
\begin{equation}
a(\vec{p},E>0)\mid k(\vec{p},E>0)\rangle=\sqrt{k(\vec{p},E>0)}\mid k(\vec{p},E>0)-1\rangle
\label{ca35}
\end{equation}
\\or the analogous ones for a negative energy single particle state
\begin{equation}
a^{\dagger}(\vec{p},E<0)\mid k(\vec{p},E>)\rangle=\sqrt{k(\vec{p},E>0)+1}\mid k(\vec{p},E>0)+1\rangle\label{ca36}
\end{equation}
\\and
\begin{equation}
a(\vec{p},E<0)\mid k(\vec{p},E>0)\rangle=\sqrt{k(\vec{p},E<0)}\mid k(\vec{p},E<0)-1\rangle
\end{equation}
\\and then extend them - formally by allowing the number
$k(\vec{p},E>0)$ of bosons in say a positive energy single particle state
with momentum $\vec{p}$ and (positive energy) to be also allowed to be
negative. You shall also allow the numbers $k(\vec{p},E<0)$ in a 
negative energy single particle state with momentum $\vec{p}$
to be both positive or zero and negative.

Then there are a couple of very important consequences:
\begin{description}
 \item[A)] You see from these stepping formulas that there is
 a ``barriere'' between the number of bosons $k$ being $k=-1$
 and $k=0$.  Operating with the annihilation operator $a$ on a
 state with $k=0$ particles give zero 
\begin{equation}
a\mid k=0\rangle=0
\end{equation}
\\and thus does not give the $\mid k=-1\rangle$ as expected from simple
stepping.  Similar one cannot with the creation operator
$a^\dagger$ cross the barriere in the opposite
direction, since
\begin{equation}
a^\dagger\mid k=-1\rangle=\sqrt{-1+1}\mid k=0\rangle=0
\end{equation}

Thus we have that the states describing the number of bosons $k$
in a given single particle state are \underline{not} \underline{connected}
by --a finite number of operations -- of creation and annihilation
operatiors.

Really this means that we make best by considering the positive
sector of the space of positive or zero number of bosons and
another sector formed from the $\mid k\rangle$ states with
$k=-1, -2, -3,...$ being a negative number of bosons.  By ordinary
creation and annihilation operators, as they would occur in
some interaction Hamiltonian, one cannot cross the barriere.
This means that if to beign with one has say a negative number of
boson in a given single particle state, then an
ordinary interaction cannot change that fact.

Thus we take it that one can choose once forever to put
some single particles states in their positive sector and
others in their negative sector, and they 
then will stay even under operation of 
an interaction Hamiltonian.
If one for example make the 
ansatz that 
all the negative
energy single particle states have a negative number of 
bosons while the positive energy states have
zero or a positive number of bosons in 
them, then this ansatz can be kept forever.

This special choice we call the ``physical choice''
and we saw already\cite{Diracsea} --and shall see very soon here --
that this choice gives us a positive definite Fock space.

\item[B)] The norm square of the states $\mid k\rangle$ (with
$k=-1, -2, \ldots $) i.e. with negative numbers 
$k$ of bosons 
have to vary alternatingly with $k$ even, $k$ odd.

Using the writing of a negative $k$
\begin{equation}
\mid k\rangle\sim\hspace{2cm} a^{|k|-1}\mid k=-1\rangle
\end{equation}

We may evaluate $\langle k\mid k\rangle\sim <-1\mid (a^\dagger)^{|k|-1}a^{|k|-1}\mid -1\rangle$
for $k\le -1$.

Now using still the usual commutation rule 
\begin{equation}
[a^\dagger,a]=-1
\end{equation}
\\you easily see that we normalize by putting
\begin{equation}
\mid k\rangle=\frac{1}{\sqrt{(|k|-1)!}}a^{|k|-1}\mid -1\rangle
\end{equation}
\\and
\begin{equation}
\langle k\mid k\rangle=(-1)^{|k|}
\label{alt}
\end{equation}
\\say for $k\le -1$ (having taken $\langle -1\mid -1\rangle=-1$.) while
of course for $k=0,1,2,\ldots$ you have $\langle k\mid k\rangle=1$.

The major success of our ``Dirac sea for bosons'' is
that one can arrange the sign alternation with (\ref{alt})
with the total number of negative energy bosons to cancel
against  the sign from in (\ref{inb}) so as to
achieve, if we choose the ``physical sector'', to get in 
total the Fock space, which has 
positive norm square.  This ``physical sector'' corresponds 
to that negative energy single particle states
are in the negative sectors, while the positive 
energy single particle states are in the positive
sector.

The basis vectors of the full Fock space for the physical
sector are thus of the form
\begin{equation}
|\ldots,k(\vec{p},E>0),\ldots;\ldots,k(\vec{p},E<0),\ldots \rangle\label{basis}
\end{equation}
\\where the dots \ldots denotes that we have one integer
number for every momentum vector --value ($\vec{p}$ or $\vec{p'}$),
but now the numbers $k(\vec{p},E>0)$ of particles in a positive
energy are-- in the physical sector-combination-- restricted to be
non-negative while the numbers of bosons in the negative energy
single particle states are restricted to be negative
\begin{eqnarray}
k(\vec{p},E>0)=&0,1,2,\ldots \nonumber
\\k(\vec{p},E<0)=&-1,-2,-3,\ldots
\end{eqnarray}

These basis vectors (\ref{basis}) are all orthogonal to
each other, and so the inner product is alone given by their
norm squares
\begin{eqnarray}
\langle\ldots, k(\vec{p},E>0),\ldots; \ldots, k(\vec{p'},E<0),\ldots \mid
\ldots, k(\vec{p},E>0),\ldots; \ldots, k(\vec{p'},E<0),\ldots \rangle\nonumber
\\=(-1)^{\sharp\mbox(neg\ energy\ b)}\prod_{\vec{p'}}(-1)^{|k(\vec{p'},E<0)|}=1\label{f216}
\end{eqnarray}
\\Here $\sharp\mbox(neg\ energy\ b)$ means the total number of negative energy bosons
i.e.
\begin{equation}
\sharp\mbox(neg\ energy\ b)=\sum_{\vec{p'}}k(\vec{p},E<0)
\end{equation}
\\(a negative number in our physical sector-combination).  
The factor $(-1)^{\sharp (neg\ energy\ b.)}$ comes from 
(\ref{inb}) which gives negative norm square for single particle
states with negative energy, because $\frac{\overleftrightarrow{\partial}}{{\partial}t}$
is essentially the energy.  The other factor $\prod_{\vec{p'}}(-1)^{|k(\vec{p'},E<0)|}$
comes from (\ref{alt}) one factor for each negative  single particle
energy state, i.e. each $\vec{p'}$. 
Had we  
here chosen another
sector-combination, e.g. to take $k(\vec{p},E<0)$ non-negative as well as
$k(\vec{p},E>0)$, then we would have instead
\begin{equation}
\left.
\begin{array}{l}
k(\vec{p},E>0)=0,1,2,\ldots
\\k(\vec{p'},E<0)=0,1,2,\ldots
\end{array}
\right\}
\mbox{(both\ pos\ sectors.)}
\end{equation}
\\and the inner with themselves, norm squares product for the
still mutually orthogonal basis vectors would be
\begin{eqnarray}
\langle\ldots, k(\vec{p},E>0),\ldots; \ldots, k(\vec{p'},E<0),\ldots \nonumber
\\\mid\ldots, k(\vec{p},E>0),\ldots; \ldots, k(\vec{p'},E<0),\ldots \rangle\nonumber
\\=(-1)^{\sharp(neg\ energy\ b)}\nonumber\\\mbox{(for\ both\ positive\ sectors)}
\end{eqnarray}
\\and that for this case (``sector combination''), the inner product is
\underline{not} positive definite.

Such strange sector combination is of course mainly of academical interest.
But for instance this last mentioned ``both positive sector'' sector-combination,
can have easily position eigenstate particles in the Fock space description.
Normally positon is not possible to be well defined in relativistic theories.


As already mentioned above, we have a slightly complicated
inner product in as far as we have sign-factors in the inner
product coming from two different sides:
\item{1)}The inner product sign-factor from the single particle
wave function coming from (\ref{inb}) gives a minus for negative
energy particles, ending up being $(-1)^{\sharp(neg.\ energy\ b)}$
in (\ref{f216}).
\item{2)}The other inner product sign factor comes from (\ref{alt}).

In the above, we have used the dagger symbol ``\hspace{0.1cm}$\dagger$\hspace{0.1cm}'' on $a^\dagger$
to denote the Hermitian conjugate w.r.t. only the inner product coming 
from (\ref{alt}), but have not included the factor from 1) meaning from
(\ref{inb}).  Thus we strictly speaking must consider also
a full dagger (full $\dagger_f$) meaning hermitian conjugation
corresponding the full inner product also including 1), i.e. the (\ref{inb})
extra minus for the negative energy states.  So although we have
not changed $a(\vec{p},E>0)$ nor $a(\vec{p},E<0)$ we have to 
distinguish two different $a^{\dagger}{'}$s namely $a^\dagger$ 
and $a^{\dagger_f}$.

In fact we obtain with this notation of two different $\dagger{(')}$s.
\begin{equation}
a^{\dagger_f}(\vec{p},E>0)=a^{\dagger}(\vec{p},E>0)
\end{equation}
\\but
\begin{equation}
a^{\dagger_f}(\vec{p},E<0)=-a^{\dagger}(\vec{p},E<0)
\end{equation}

Since at the end, the physical/usual second quantized
Boson-theory has as its inner product the full
inner product one should, in the physical use, use the
Hermitian conjugation $\dagger_f$.  So the creation
operators to be identified with creation operators
are respectively:
\\For a \underline{particle};
\begin{equation}
a^{\dagger}_{usual}(\vec{p})=a^{{\dagger}f}(\vec{p},E>0) 
\end{equation}
\\while for an \underline{antiparticle} of momentum $\vec{p}$ it is;
\begin{equation}
a^{\dagger}_{usual\ anti}(\vec{p})=a(-\vec{p},E<0) 
\end{equation}

Similarly:
\begin{eqnarray}
a_{usual}(\vec{p})=a(\vec{p},E>0)\nonumber
\\a_{usual\ anti}(\vec{p})= a^{{\dagger}_f}(-\vec{p},E<0)=-a^{\dagger}(-\vec{p},E<0)
\end{eqnarray}

Using the extended commutation rules
\begin{equation}
[a(\vec{p},\gtrless E),a^{\dagger}(\vec{p'},\gtrless E)]=\delta_{\vec{p'}\vec{p}}\cdot
\begin{cases}
1\ \mbox{for\ same $<$ or $>$}
\\0\ \mbox{for\ different $<$ or $>$}
\end{cases}
\end{equation}
\\so that for instance
\begin{equation}
[a(\vec{p},<E),a^{\dagger}(\vec{p'},<E)]=\delta_{\vec{p}\vec{p'}}
\end{equation}

We quickly derive  the correspondingcommutation rules using the
``full dagger''
\begin{equation}
[a(\vec{p},E>0),a^{{\dagger}_f}(\vec{p'},E>0)]=\delta_{\vec{p}\vec{p'}}
\end{equation}
\\and
\begin{equation}
[a(\vec{p},E<0),a^{{\dagger}_f}(\vec{p'},E<0)]=\delta_{\vec{p}\vec{p'}}
\end{equation}
 
\end{description}

\section{``Majorana-bosons''}
\label{Sec 4}


We shall now in this section analogously to what we did
in sections 2 for Fermions as a warm up excercise from our
Fock space defined in section 3 for e.g. the physical sector-
combination extract a subspace $H_{Maj}$ and on that find a
description of now bosons which are their own antiparticles.
There would be some meaning in analogy to the Fermion case
to call such bosons which are their own antiparticles
by the nickname ``Majorana-bosons''.

As for the fermions we shall expect a state with say $k_{Maj}(\vec{p})$
``Majorana-bosons''with momentum equal to $\vec{p}$ to be 
presented as a superposition of a number of the
``essentially charged'' bosons or antibosons of the
type discussed in foregoing section.  Here an antiparticle
of course means that one has made the number of bosons
in a negative energy single particle state one unit more negative.
Typically since the physical vacuum has $k(\vec{p},E<0)=-1$ for all
momenta and an antiparticle of momentum $\vec{p}$ would mean 
that $k(-\vec{p},E<0)$ gets decreased from $-1$ to $-2$.  If you 
have several antiparticles $l$ say in the same state with
momentum $\vec{p}$ of course you decrease
$k(-\vec{p},E<0)$ to $-1-l,k(-\vec{p},E<0)=-1-l$ (for $l$ antiparticles).

In other words we expect a state with say $l_{Maj}$ ``Majorana-bosons'' with
momentum $\vec{p}$ to be a superposition of states in the Fock space with 
the number of antiparticles running from $l=0$ to $l=l_{Maj}$ while 
correspondingly the number with momentum $\vec{p}$ is made to
$l_{Maj}-l$ so that there are together in the representing state just
equally many particles or antiparticles as the number of
``Majorana-bosons'' $l_{Maj}$ wanted.

We actually hope --and we shall see we shall succeed-- that we 
can construct a ``Majorana-boson'' creation operator for say
a ``Majorana-boson'' with momentum $\vec{p}$,$b^{\dagger}(\vec{p})$
analogously to the expressions (\ref{cref}) and (\ref{anf})
\\$b^{\dagger}(\vec{p})=\frac{1}{\sqrt{2}}\left(a^{\dagger}(\vec{p},E>0)+a(-\vec{p},E<0)\right)$
and $b(\vec{p})=\frac{1}{\sqrt{2}}\left(a(E>0)+a^{\dagger}(E<0)\right)$.

Since an extra phase on the basis states does not matter so much we could
also choose for the boson the ``Majorana boson'' creation and annihilation
operators to be 
\begin{align}
b^{\dagger_f}(\vec{p})&=\frac{1}{\sqrt{2}}\left(a^{\dagger}(E>0)+a(-\vec{p},E<0)\right)\nonumber\label{f228}
\\&=\frac{1}{\sqrt{2}}\left(a^{\dagger_f}(\vec{p},E>0)+a(-\vec{p},E<0)\right)\nonumber
\\b(\vec{p})&=\frac{1}{\sqrt{2}}\left(a(\vec{p},E>0)+a^{\dagger_f}(-\vec{p},E<0)\right)\nonumber
\\&=\frac{1}{\sqrt{2}}\left(a(\vec{p},E>0)-a^{\dagger}(-\vec{p},E<0)\right).
\end{align}

One must of course then check --first on the physical sector-combination
but later on others-- that $b^{\dagger}(\vec{p})$ and $b(\vec{p})$ obey
the usual commutation rules
\begin{align}
\left[b(\vec{p}),b(\vec{p'})\right]&=0 \nonumber
\\\left[b^{\dagger_f}(\vec{p}),b^{\dagger_f}(\vec{p'})\right]&=0 \nonumber
\\\left[b(\vec{p}),b^{\dagger_f}(\vec{p'})\right]&=\delta _{\vec{p}\vec{p'}}.
\end{align}

We also have to have a vacuum for the ``Majorana-boson'' theory,
but for that we use in the physical sector-combination theory the
same state in the Fock space as the one for the ``essentially charged''
boson system.  This common physical vacuum state (in the Fock space) is
characterized as the basis vector
\begin{equation}
\mid \ldots, k(\vec{p},E>0),\ldots;\ldots,k(\vec{p},E>0),\ldots\rangle
\end{equation}
with
\begin{equation}
k(\vec{p},E>0)=0\  \mbox{for\ all\ $\vec{p}$}
\end{equation}
and
\begin{equation}
k(\vec{p},E<0)=-1\  \mbox{for\ all\ $\vec{p}$}
\end{equation}
Indeed we also can check then of course that defining
\begin{equation}
\mid vac\ phys\rangle =\mid \ldots, k(\vec{p},E>0)=0,\ldots;\ldots,k(\vec{p'},E<0)=-1,\ldots \rangle
\end{equation}
we have
\begin{equation}
b(\vec{p})\ \mid vac\ phys\rangle=0
\end{equation}

On the other hand, we can also see that e.g.
\begin{align}
&\frac{1}{\sqrt{l_{Maj}!}}\left(b^{\dagger_f}(\vec{p})\right)^{l_Maj} \mid vac\ phys\rangle\nonumber\label{expl.}
\\
&=\frac{1}{2^{l_{Maj}/2}} \cdot\Sigma_{l}\begin{pmatrix} l_{Maj}\\l \end{pmatrix}
\left(a^{\dagger}(\vec{p},E>0)\right)^{l}a(-\vec{p},E<0)^{l_{Maj}-l} \mid vac\ phys \rangle\nonumber
\\
&=\Sigma_{l}\begin{pmatrix}{l}_{Maj}\\l \end{pmatrix}\ \mid\ldots,k(\vec{p},E>0)=l,\ldots;\ldots,k(-\vec{p},E<0)=l_{Maj}-l,\ldots\rangle \nonumber
\\&\cdot\sqrt{l!}\sqrt{(l_{Maj}-l)!}\cdot\frac{1}{\sqrt{l_{Maj}!}}\nonumber
\\
&=\frac{1}{2^{l_{Maj}/2}} \cdot\Sigma_{l}\sqrt{\begin{pmatrix} l_{Maj}\\l \end{pmatrix}} \mid\ldots,k(\vec{p},E>0),\ldots;
\ldots,k(-\vec{p},E<0),\ldots\rangle
\end{align}

If we only put the Majorana-boson  particles into the momentum $\vec{p}$ state of course only
$k(\vec{p},E>o)$ and $k(-\vec{p},E<0)$ will be different from their $\mid phys\ vac\rangle$
values $0$ and $-1$ respectively for $E>0$ and $E<0$.  But really the extension to put 
``Majorana-bosons'' in any number of momentum states is trivial.

We now have also to construct the analogous operator to the $r(\vec{p})$ for the
fermions so that we can characterize the subspace $H_{Maj}$ to be for the boson case
\begin{equation}
H_{Maj}=\left\{\mid \rangle \mid r(\vec{p}) \mid \rangle=0\right\}.
\end{equation}

We in fact shall see that the proposal
\begin{eqnarray}
r(\vec{p})&=&\frac{1}{\sqrt{2}}\left(a(+\vec{p},E>0)+a{\dagger}(-\vec{p},E<0)\right)\nonumber
\\&=&\frac{1}{\sqrt{2}}(a(\vec{p},E>0)-a^{{\dagger}_f}(-\vec{p},E<0)
\end{eqnarray}
\\does the job.
\\Now we check (using (\ref{f228}))
\begin{align}
\left[r(\vec{p}),b^{\dagger_f}(\vec{p})\right]&=
\left[\frac{1}{\sqrt{2}}\left(a(\vec{p}),E>0)+a^{\dagger}(-\vec{p},E<0)\right),
\frac{1}{\sqrt{2}}\left(a^{\dagger}(\vec{p}),E>0)+a(-\vec{p},E<0)\right)\right] \nonumber
\\ \mbox{or}\nonumber
\\&=\frac{1}{2}\left[\left(a(\vec{p},E>0)-a^{{\dagger}_f}(-\vec{p},E<0)),
(a^{{\dagger}_f}(\vec{p},E>0)+a(-\vec{p},E<0)\right)\right] \nonumber
\\&=\frac{1}{2}(1-1)=0
\end{align}
\\and also
\begin{align}
\left[r(\vec{p}),b(\vec{p})\right]&=\left[\frac{1}{\sqrt{2}}\left(a(\vec{p}),E>0)-a^{\dagger_f}(-\vec{p},E<0)\right),
\frac{1}{\sqrt{2}}\left(a(\vec{p}),E>0)+a^{\dagger_f}(-\vec{p},E<0)\right)\right]\nonumber
\\&=\left[\frac{1}{\sqrt{2}}\left(a(\vec{p}),E>0)+a^{\dagger}(-\vec{p},E<0)\right),
\frac{1}{\sqrt{2}}\left(a(\vec{p}),E>0)-a^{\dagger}(-\vec{p},E<0)\right)\right]\nonumber
\\&=0
\end{align}

We should also check that the physical vacuum
\begin{equation}
\mid \ phys\ vac\rangle =\mid \ldots, k(\vec{p},E>0)=0,\ldots;\ldots,k(\vec{p'},E<0)=-1,\ldots \rangle
\end{equation}
\\in which there is in all negative energy (with momentum $\vec{p'}$ say)
single particle states $k(\vec{p'},E<0)=-1$ bosons, and in all
positive energy single particle states $k(\vec{p},E>0)=0$ bosons is annihilated
by the $r(\vec{p})$ operators.

Now indeed
\begin{align}
&r(\vec{p})\mid \ phys\ vac\rangle\nonumber 
\\&=\frac{1}{2}\left(a(\vec{p},E>0)-a^{\dagger_f}(-\vec{p},E,0)\right)\mid phys\ vac\rangle\nonumber
\\&=\frac{1}{2}\left(a(\vec{p},E>0)+a^{\dagger}(-\vec{p},E,0)\right)\cdot \mid \ldots, 
k(\vec{p},E>0)=0,\ldots;\ldots,k(\vec{p},E<0)\nonumber
\\&=-1,\ldots \rangle\nonumber
\\&=0
\end{align}
\\basically because of the barriere, meaning the square roots in the 
formulas (\ref{ca35},\ref{ca36})
became zero.


The result of this physical section for the most
attractive formalism with $b(\vec{p})$
and 
$b^{{\dagger_f}}(\vec{p})$ annihilating and creating
operators for the Boson-type particle being its own antiparticle (=Majorana Boson)
and to them corresponding useful state condition operator
$r_{f}(\vec{p})$ is summarized as:
\begin{align}
b(\vec{p})=\frac{1}{\sqrt{2}}\left(a(\vec{p},E>0)-a^{{\dagger_f}}(-\vec{p},E<0)\right)\nonumber
\\b^{{\dagger_f}}(\vec{p})=\frac{1}{\sqrt{2}}\left(a^{{\dagger_f}}(\vec{p},E>0)-a(-\vec{p},E<0)\right)\nonumber
\\ \mid phys\ vac \rangle=\mid \ldots, k(\mbox{all}\ \vec{p},E>0)=0,\ldots;\ldots, k(\mbox{all}\ \vec{p},E<0)=-1,\ldots\rangle \nonumber
\\r_{f}(\vec{p})=\frac{1}{\sqrt{2}}\left(a(\vec{p},E>0)+a^{{\dagger_f}}(-\vec{p},E<0)\right)
\end{align}
\\the useful subspace for bosons being their own antiparticles being
\begin{equation}
H_{Maj\ f}=\left\{\mid \ \rangle\ \mid \forall_{\vec{p}}r_{f}(\vec{p}) \mid \ \rangle=0\right\}
\end{equation}
\\(One should note that whether one chooses our $r(\vec{p'})'s$ or the $r_{f}(\vec{p})'s$ to
define makes 
no difference for the space $H_{Maj\ f}$ rather than $H_f$, 
since we actually have $r(\vec{p}) = r_f
(\vec{p})$ the two expressions being just 
expressed in terms of different $a^\dagger(\vec{p}, E<0)$ and $a^{\dagger_f}(\vec{p}, E<0)$ say)

We can easily check that our explicit state expressions (\ref{expl.}) indeed are 
annihilated
by $r(\vec{p})$ 
It were formally left out the$E>0$ or $E<0$ for the 
$b(\vec{p})$ and $b^{{\dagger_f}}(\vec{p})$ it being understood that $E>0$, but formally
we can extrapolate also to $E<0$ and it turns of $b(\vec{p},E>0)=b^{{\dagger_f}}(-\vec{p},E<0)$?

\subsection{Charge Conjugation Operation}
Since we discuss so much bosons being their own antiparticles coming out of
a formalism in which the bosons --at first-- have antiparticles \underline{different} from 
themselves, we should here define a charge conjugation operator ${\bf C}$ that transform a boson
into its antiparticle:
\\That is to say we want this operator acting on the Fock space to have the commutation
properties with our creation and annihilation operators
\begin{equation}
{\bf C}^{-1}a(\vec{p},E>0){\bf C}=a^{{\dagger_f}}(-\vec{p},E<0)
\end{equation}
\\and
\begin{equation}
{\bf C}^{-1}a^{{\dagger_f}}(\vec{p},E>0){\bf C}=a(-\vec{p},E<0).
\end{equation}
\\We also have
\begin{equation}
{\bf C}^{-1}a(\vec{p},E<0){\bf C}=a^{{\dagger_f}}(-\vec{p},E>0)
\end{equation}
\\and
\begin{equation}
{\bf C}^{-1}a^{{\dagger_f}}(\vec{p},E<0){\bf C}=a(-\vec{p},E>0)
\end{equation}

These requirements suggest that we on the basis of (\ref{basis})
for the Fock space have the operation 
\begin{align}
{\bf C}\mid& \ldots \stackrel{\sim}{k}(\vec{p},E>0),\ldots;\ldots,\stackrel{\sim}{k}(\vec{p'},E<0),\ldots\rangle\nonumber\label{cdef}
\\=\mid&\ldots,k(\vec{p},E>0)=-\stackrel{\sim}{k}(-\vec{p'},E<0)+1,\ldots;
\\
&\ldots,k(\vec{p'},E<0)\nonumber
=-1-\stackrel{\sim}{k}(-\vec{p'},E>0),\ldots\rangle.
\end{align}

Using the ``full inner product'' this ${\bf C}$ operation conserves the norm,
and in fact it is unitary under the full inner product corresponding 
hermitean conjugation ${\dagger_f}$ i.e.
\begin{equation}
{\bf C}^{{\dagger_f}}{\bf C}={\bf 1}={\bf C}{\bf C}^{{\dagger_f}}
\end{equation}

But if we used the not full inner product, so that the norm squares for
basis vector would be given by (\ref{ns}) and therefore corresponding hermitean
conjugation $\dagger$, then if ${\bf C}$ acts on a state in which the difference
of the number of positive and negative energy bosons is odd, the
norm square would change sign under the operation with ${\bf C}$.

So under $\dagger$ the charge conjugation operator could not 
possibly be unitary:
\begin{equation}
{\bf C}^{\dagger}{\bf C}\neq {\bf 1} \neq {\bf C}{\bf C}^\dagger
\end{equation}

\section{``Majorana boson'' in unphysical sector-combination}


As an example of one of the unphysical sector-combination
we could take what in our earlier work ``Dirac sea for Bosons''
were said to be based on the naive vacuum.  This naive vacuum theory
means a theory in which we do not make any emptying vacuum
but rather let there be in both positive and negative single particle 
energy states a positive or zero number of particles.  So in this naive
vacuum attached sector combination we can completely ignore the
extrapolated negative number of boson possibilities; we so to 
speak could use the analogue of the harmonic oscillator with 
normalized states only.

This means that the inner product excluding the
negative single particle state normalization using
(\ref{inb}) will be for this naive vacuum sector combination
completely positive definite.

However, including the negative norm factor for the negative
energy states from (\ref{inb}) so as to get the full inner
product we do no longer have the positive definite Hilbert
inner product on the Fock space.  Now rather we have for
basis vectors (\ref{basis}) instead of (\ref{f216}) that the
norm squares 
\begin{eqnarray}
&\langle\ldots,k(\vec{p},E>0),\ldots;\ldots,k(\vec{p'},E<0),\ldots\mid\label{ns}
\ldots,k(\vec{p},E>0),\ldots;\ldots,k(\vec{p'},E<0),\ldots\rangle\nonumber
\\&=(-1)^{\sharp (neg.\ energy\ b.)}
\end{eqnarray}

This means that the norm square of a basis vector is positive when 
the number of negative energy bosons is even, but negative when the 
number of negative energy bosons is odd.

In the naive vacuum sector combination the vacuum analogue Fock
space state is the ``naive vacuum'',
\begin{equation}
\mid\ naive\ vac.\rangle=\mid\ldots,k(\vec{p},E>0)=0,\ldots;\ldots,k(\vec{p},E<0)=0,\ldots\rangle.\label{f259}
\end{equation}

In analogy to what we did in the foregoing section 4, we should then construct the
states with various numbers of bosons of the Majorana type being their own antiparticles
by means of some creation and annihilation operators $b^{\dagger_f}(\vec{p})$ and
$b(\vec{p})$, but first one needs a vacuum that is its own ``anti state'' so to speak,
meaning that the charge conjugation operator ${\bf C}$ acting on it gives it back.
i.e. one need a vacuum $\mid vac?\rangle$ so that 
\begin{equation}
{\bf C}\mid vac?\rangle=\mid vac?\rangle
\end{equation}

But this is a trouble!  The ``naive vacuum'' $\mid naive\ vac.\rangle$ in not left 
invariant under the charge conjugation operator ${\bf C}$ defined in the last subsection of 
Section 4 by (\ref{cdef}).

Rather this naive vacuum is by ${\bf C}$ transformed into a quite different sector
combination, namely in that sector combination, in which there is
a negative number of bosons in both positive and negative energy single particle 
eigenstates.  i.e. the charge conjugation operates between one 
sector combination and another one!  But this then means, that we
cannot make a representation of a theory with (only) bosons being
their own antiparticles unless we use more than just the naive
vacuum sector combination.  i.e. we must include also the
both number of particles being negative sector combination.

In spite of this need for having the two sector combinations --both the
naive all positive particle number and the opposite all negative
numbers of particles-- in order that the charge conjugation 
operator should stay inside the system --Fock space,  we should still have in 
mind that the 
creation and annihilation 
\underline{operators}
cannot pass the barriers and thus can not go from sectors, also
the inner product between different sector combinations are
divergent and ill defined (and we should either avoid such inner products
or define them arbitrarily).

So if we construct ``Majorana boson'' creation and annihilation operators
analogoulsy to the $b(\vec{p})$ and $b^{\dagger_f}(\vec{p})$ in foregoing
section as a linear combination of $a^{(\dagger)}(\vec{p},\gtrless E)$ operators
operating with such $b(\vec{p})$ and $b^{\dagger_f}(\vec{p'})s$ will stay 
inside one sector combination.  For instance such $b(\vec{p})$ and 
$b^{\dagger_f}(\vec{p})$ constructed analogously to the physical sector ones
formally would operate arround staying inside the naive vacuum sector combination
if one starts there, e.g. on the naive vacuum $\mid naive\ vac.\rangle$.  In this
--slightly cheating way-- we could then effectively build up a formalism for
bosons which are their own antiparticles inside just \underline{one} sector combination.
When we say that it is ``slightly cheating'' to make this construction on only
one sector combination it is because we cannot have the true antiparticles if
we keep to a sector combination only, which is not mapped into itself by
the charge conjugation operator $\bf{C}$.  It namely then would mean that the
true antiparticle cannot be in the same sector combination.

Nevertheless let us in this section 5 study precisely this ``slightly
cheating'' formalism of keeping to the naive vacuum sector combination with
positive numbers of particles only.

We then after all simply use the naive vacuum $\mid naive\ vac.\rangle$ defined by
(\ref{f259}) as the ``Majorana boson''-vacuum although it is not invariant under
$\bf{C}$, which we must ignore or redefine, if this shall be o.k.

We may e.g. build up a formalism for the slightly cheating Majorana bosons 
by starting from the $\mid naive\ vac.\rangle$ (\ref{f259}) and build up with
$b^{\dagger_f}(\vec{p})$ taken to be the same as (\ref{f238})
\begin{equation}
b^{\dagger_f}(\vec{p})=\frac{1}{\sqrt{2}}\left(a^{\dagger_f}(\vec{p},E>0)+a(-\vec{p},E<0)\right)
\end{equation}
\\and
\begin{align}
b(\vec{p})&=\frac{1}{\sqrt{2}}\left(a(\vec{p},E>0)+a^{\dagger_f}(-\vec{p},E<0)\right)\nonumber
\\&=\frac{1}{\sqrt{2}}\left(a(\vec{p},E>0)-a^{\dagger}(-\vec{p},E<0)\right)
\end{align}

We have already checked that for all sector combinations we have
\begin{equation}
\left[b(\vec{p},b^{\dagger_f}(\vec{p'})\right]=\delta _{\vec{p}\vec{p'}} \label{f280a}
\end{equation}
\\and of course
\begin{align}
\left[b(\vec{p},b(\vec{p'}\right]&=0 \nonumber
\\&=\left[b^{\dagger_f}(\vec{p}),b^{\dagger_f}(\vec{p'})\right] \label{f280b}
\end{align}

So we see that we can build up using $b(\vec{p})$ and $b^{\dagger_f}(\vec{p})$
a tower of states with any nonnegative number of what we can call the
Majorana bosons for any momentum $\vec{p}$.

We can also in all the sector combinations use the already constructed
\begin{align}
r(\vec{p})&=\frac{1}{2}\left(a(\vec{p},E>0)-a^{\dagger_f}(-\vec{p},E<0)\right)\nonumber\label{f281}
\\&=\frac{1}{2}\left(a(\vec{p},E>0)+a^{\dagger}(-\vec{p},E<0)\right)
\end{align}
\\to fullfill the commutation conditions
\begin{align}
\left[r(\vec{p}),b^{\dagger_f}(\vec{p})\right]&=0\nonumber
\\\left[r(\vec{p}),b(\vec{p'})\right]&=0
\end{align}
\\and we even have
\begin{equation}
r(\vec{p})\mid naive\ vac.\rangle=0
\end{equation}

So indeed we have gotten a seemingly full theory of ``Majorana Bosons'' inside the 
naive vacuum sector combination subspace
\begin{equation}
H_{Maj}=\{\mid \rangle \mid \forall_{\vec{p}}\left(r(\vec{p}) \mid \rangle=0\right)\}
\end{equation}
\\but it is not kept under the $\bf{C}$ as expected.

But really what we ended up constructing were only a system of positive energy particle
states since the creation with $b^{\dagger_f}(\vec{p})=b^{\dagger}(\vec{p})$ starting from
the naive vacuum only produces positive energy particles in as far as the
$a(-\vec{p},E<0)$ contained in $b^{\dagger_f}(\vec{p})$ just gives zero on the naive
vacuum.

So this a ``bit cheating'' formalism really just presented for us the ``essentially
charged'' positive energy particles as ``the Majorana-bosons''.

That is to say this a bit cheating formalism suggests us to use in the naive vacuum sector combination
the ``essentially charged particles'' as were they their
own antiparticles.

If we similarly built a Majorana boson Fock space system of the 
\begin{align}
{\bf{C}} \mid naive\ vac.\rangle&=\mid vac.\ with\ both\ E>0\ and\ E<0\ emptied\ out\rangle\nonumber
\\&=\mid\ldots,k(\vec{p},E>0)=-1,\ldots;\ldots,k(\vec{p},E<0)=-1,\ldots\rangle,
\end{align}
we would obtain a series of essentially antiparticles (with positive energies) constructed
in the ``both numbers of bosons negative'' sector combination.

What we truly should have done were to start from the superposition
\begin{align}
\mid self\ copy\ vac.\rangle&\hat{=}\frac{1}{\sqrt{2}}\left(\mid naive\ vac.\rangle+{\bf{C}}\mid naive\ vac.\rangle\right)\nonumber
\\&=\frac{1}{\sqrt{2}}\left(\mid\ldots,k(\vec{p},E>0)=0,\ldots;\ldots,k(\vec{p},E<0)
=0,\ldots \rangle 
\right.
\nonumber \\
& \quad
\left.
+\mid\ldots,k(\vec{p},E>0)-1,\ldots;\ldots,k(\vec{p},E<0)=-1,\ldots\rangle \right)
\end{align}
\\and then as we would successively go up the latter with $b^{\dagger_f}(\vec{p})$ operators
we would successively fill equally many positive energy particles into the
$\mid naive\ vac.\rangle$ and positive energy antiparticles in
${\bf{C}}\mid naive\ vac.\rangle$.
Note that analogously to the above called
``a bit cheating'' Majorana-boson 
construction using {\em only the positive 
energy single particle states} we obtain 
here only use of the positive energy states for the naive vacuum sector combination 
and only the negative energy single particle states for the Charge conjugation to the 
naive vacuum sector combination. Also it should not be misunderstood: The filling 
in is {\em not} running parallel in the 
sense that the sectors truly follow each other. Rather one has to look for if there is Majorana boson by looking into {\em both}
sector-combination- projections.

So we see that what is the true Majorana boson theory built on the two unphysical
sector combinations having respectively nonzero numbers of particles (the naive vacuum construction)
and negative particles number in both positive and negative energies is the following:

A basis state with $n(\vec{p})$ Majorana bosons with momentum $\vec{p}$, -and as we always have for
Majorana's positive energy- gets described as a superposition ot two states --one from
each of the two sector combinations-- with just $n(\vec{p})$ ordinary 
(positive energy) essentially charged bosons (of the original types of our
construction created by $a^{\dagger}$..) and a corresponding Fock space state from the other
sector, now with $n(\vec{p})$ antiparticles in the other sector combination
(the one built from ${\bf{C}}\mid naive\ vac.\rangle$).

Both of these separate sector combinations have for the used states 
a positive definite Hilbert space.

As already stated the overlap between different sector combinaions vectors are
divergent and illdefined.

We can check this rather simple way of getting the Majorana bosons described in our on the state
$\frac{1}{2}\left(\mid naive\ vac.\rangle + {\bf{C}}\mid naive\ vac.\rangle\right)$
built system of states by noting what the condition $r(\vec{p})\mid\rangle=0$ tells us
the two sector combinations:

On a linear combination of basis vectors of the naive vacuum construction type 
\begin{eqnarray}
\mid\rangle=\Sigma 
\mid\ k(\vec{p},E>0)\ge 0,\ldots;\ldots,k(\vec{p},E<0)\ge 0,\ldots\rangle\nonumber
\\C_{\ldots k(\vec{p},E>0)\ldots;\ldots\tilde{k}(\dot{\vec{p}},E<0)\ldots}
\end{eqnarray}
\\the requirement
\begin{equation}
r(\vec{p})\mid\rangle=0
\end{equation}
\\relates coefficients which correspond to basis states being connected by
$k(\vec{p},E>0)$ going one up while $k(-\vec{p},E<0)$ going one unit down or opposite.
As we get the relation
\begin{eqnarray}
\sqrt{1+k(\vec{p},E>0)}C_{\ldots k(\vec{p},E>0)+1,\ldots;\ldots,k(-\vec{p},E<0),\ldots} \nonumber
\\+C_{\ldots,k(\vec{p'},E>0),\ldots;\ldots,
k(-\vec{p},E<0)-1,\ldots\cdot}\sqrt{k(-\vec{p},E<0)}=0
\end{eqnarray}
\\we can easily see that the states being annihilated are of the form
\begin{equation}
\sum_{\tiny\begin{matrix}k(\vec{p},E>0)\\ and\ the\ difference\\d=k(\vec{p},E>0)-k(-\vec{p},E<0)\\ FIXED.\end{matrix}}
 (-1)^{k(\vec{p},E>0)}\frac{\sqrt{k(-\vec{p},E<0)!}}{\sqrt{k(\vec{p},E>0)!}}\cdot
 \mid\ldots,k(\vec{p},E>0),\ldots;\ldots,k(-\vec{p},E<0),\ldots\rangle.\label{f292}
\end{equation}

As a special case we might look at possibility that the difference 
\begin{equation}
d=k(\vec{p},E>0)-k(-\vec{p},E<0)
\end {equation}
\\were 0.  In this case the $\mid naive\ vac.\rangle$ itself would
be in the series.  In this case the solution (\ref{f292}) reduces to
\begin{equation}
\sum_{k=0}(-1)^k\mid\ldots,k(\vec{p},E>0)=k,\ldots;\ldots,k(-\vec{p},E<0)=k,\ldots\rangle
\end{equation}

But it is now the problem that this series does not converge.  But for appropriate
values of the difference $d$,
\begin{equation}
d\ge 2,
\end{equation}
\\the series (\ref{f292})converge.

For the convergent cases we can estimate the norm square of a state (\ref{f292})
to go proportional to
\begin{equation}
\| \mid\rangle \|^{2}\propto\sum_{k=0}^{\infty} \frac{(k-d)!}{k!}(-1)^{k-d}
\end{equation}
\\where the $(-1)^{k-d}$ now comes from the alternating ``full'' norm square due to the
factor $(-1)^{\sharp (neg.\ energy\ b.)}$.  This expression in turn is proportional to
\begin{align}
\sum_{k=0}^{\infty}\begin{pmatrix}k-d\\-d\end{pmatrix}(-1)^{k-d}
&=\sum_{n=-d}\begin{pmatrix}n\\-d\end{pmatrix}(-1)^{n}
\qquad (n=k-d)
\nonumber
\\&=\frac{(-1)^{-d}}{\left(1-(-1)\right)^{-d+1}}
\end{align}
\\which is zero for $d-1\ge 1$.

So indeed it is seen that the basis states in $H_{Maj}$ part inside the naive
vacuum sector combination has zero norm.  Since the states with different
numbers of Majorana-bosons are represented by mutually orthogonal it means that
the whole part of the naive vacuum sector combination used to represent the
Majorana-bosons has totally zero inner product.  Basically that means that the inner 
product transfered from the original theory with its ``essentially charged
bosons'' to the for Majorana bosons in subspace $H_{Maj}$ turns out to be zero.

This result means --extrapolating to suppose zero norm also in the divergent
cases-- that in the unphysical sector combination we get no non-trivial inner
product for the Majorana-bosons.

If ones use the true Majorana boson description by as necessary combining two
sector combinations, one could use the ambiguity (and divergence) of the 
inner product of states from different sectors to make up instead a non trivial
inner product.

\subsection{Overview of All four Sector Combinations}
Strictly speaking we could make an 
infinite number of sector combinations,
because we for every single particle state
-- meaning for every combination of a 
spin state and a momentum say $\vec{p}$ -- 
could choose for just that single particle
state to postulate the second quantized 
system considered to be started at such 
a side of the ``barriers'' that just this 
special single particle state had always a
negative number of bosons in it. For
another one we could instead choose to
have only a non-negative numbe of bosons.
Using all the choice possibilities of 
this type would lead us so to speak to 
the infinite number of 
sector combinations 
$2^{\# \hbox{``single particle states''}}$, where $\#
\hbox{``single particle states''}$ means the 
number of single particle states. But most
of these enormously many 
sector combinations would not be Lorentz 
invariant nor rotational invariant. Really,
since the sector combination should 
presumably rather be considered a part 
of the initial state condition than of 
the laws of Nature, it might be o.k. that
it be  not Lorentz nor rotational 
invariant. Nevertheless we strongly 
suspect that it is the most important to
consider the Lorentz and rotational 
invariant sector-combination-choices.
Restricting to the latter we can only 
choose a seprate sector for the positive 
enegry states and for the negative energy
sector, and then there 
would be only $2^{2}=4$ 
sector combinations.

Quite generally we have the usual rules 
for creation and annihilation
operators,but you have to have in mind 
that we have two different hermitean
conjugations denoted respectively by $\dagger$ and by $\dagger_f$, and that the 
creation operators
constructed from the same annihilation 
operators are related
\begin{align}
a^{\dagger_f}(\vec{p},E>0)&=a^{\dagger}(\vec{p},E>0)\nonumber
\\a^{\dagger_f}(\vec{p},E<0)&=-a^{\dagger}(\vec{p},E<0)
\end{align}
\\These ``usual'' relations are
\begin{align} 
\left[a(\vec{p},E>0),a^{\dagger}(\vec{p'},E>0)\right]&=\delta_{\vec{p}\vec{p'}}\nonumber
\\\left[a(\vec{p},E<0),a^{\dagger}(\vec{p'},E<0)\right]&=\delta_{\vec{p}\vec{p'}}\nonumber
\\\left[a(\vec{p},E>0),a^{\dagger_f}(\vec{p'},E>0)\right]&=\delta_{\vec{p}\vec{p'}}\nonumber
\\\left[a(\vec{p},E<0),a^{\dagger_f}(\vec{p'},E<0)\right]&=-\delta_{\vec{p}\vec{p'}}
\end{align}
\\while we have exact commutation for $a$ with $a$ or for $a^{\dagger}$ or
$a^{\dagger_f}$ with $a^{\dagger}$ or $a^{\dagger_f}$.  Each $a(\vec{p},E\gtrless0)$ or 
$a^{\dagger_f}$ or $a^{\dagger}$ act changing only the number of particle
in just the single relevant single particle state, meaning it changes only
$k(\vec{p},E\gtrless0)$; the rules are as seen analytical continuations generally
\begin{align}
&a^{\dagger}(\vec{p},E>0)\mid\ldots,k(\vec{p},E>0),\ldots;\ldots,k(\vec{p'},E<0),\ldots\rangle\nonumber
\\
=&\sqrt{k(\vec{p},E>0)+1}\mid\ldots,k(\vec{p},E>0)+1,\ldots;\ldots,k(\vec{p'},E<0),\ldots\rangle\nonumber
\\
&a^{\dagger}(\vec{p'},E<0)\mid\ldots,k(\vec{p},E>0),\ldots;\ldots,k(\vec{p'},E<0),\ldots\rangle\nonumber
\\=&\sqrt{k(\vec{p'},E<0)+1}\mid\ldots,k(\vec{p},E>0),\ldots;\ldots,k(\vec{p'},E<0)+1,\ldots\rangle\nonumber
\\&a^{\dagger_f}(\vec{p},E>0)\mid\ldots,k(\vec{p},E>0),\ldots;\ldots,k(\vec{p'},E<0),\ldots\rangle\nonumber
\\=&\sqrt{k(\vec{p},E>0)+1}\mid\ldots,k(\vec{p},E>0)+1,\ldots;\ldots,k(\vec{p'},E<0),\ldots\rangle\nonumber
\\&a^{\dagger_f}(\vec{p'},E<0)\mid\ldots,k(\vec{p},E>0),\ldots;\ldots,k(\vec{p'},E<0),\ldots\rangle\nonumber
\\=&-\sqrt{k(\vec{p'},E<0)+1}\mid\ldots,k(\vec{p},E>0),\ldots;\ldots,k(\vec{p'},E<0)+1,\ldots\rangle\nonumber
\\&a(\vec{p},E>0)\mid\ldots,k(\vec{p},E>0),\ldots;\ldots,k(\vec{p'},E<0),\ldots\rangle\nonumber
\\=&\sqrt{k(\vec{p},E>0)}\mid\ldots,k(\vec{p},E>0)-1,\ldots;\ldots,k(\vec{p'},E<0),\ldots\rangle\nonumber
\\&a(\vec{p'},E<0)\mid\ldots,k(\vec{p},E>0),\ldots;\ldots,k(\vec{p'},E<0),\ldots\rangle\nonumber
\\=&\sqrt{k(\vec{p'},E<0)}\mid\ldots,k(\vec{p},E>0)+1,\ldots;\ldots,k(\vec{p'},E<0)-1,\ldots\rangle
\end{align}

The four sector combination with the same sector for the same sign of the energy E of the single particle 
states were called:
\begin{description}
  \item {1)}The ``physical sector'' has 
  \begin{align}
  k(\vec{p},E>0)&=0,1,2,\ldots\nonumber
  \\k(\vec{p},E<0)&=-1,-2,-3\ldots
  \end{align}
  \item{2)}The ``sector-combination constructed from the naive vacuum'' has
  \begin{align}
   k(\vec{p},E>0)&=0,1,2,\ldots\nonumber
   \\ k(\vec{p},E<0)&=0,1,2,\ldots
  \end{align}
  \item{3)}The ``both sectors with negative numbers'' sector-combination has
  \begin{align}
  k(\vec{p},E>0)&=-1,-2,-3\ldots\nonumber
  \\k(\vec{p},E<0)&=-1,-2,-3\ldots
  \end{align}
  \item{4)}The ``a positive number with negative energy and vise versa'' has
  \begin{align}
  k(\vec{p},E>0)&=-1,-2,-3\ldots\nonumber
  \\ k(\vec{p},E<0)&=0,1,2,\ldots
  \end{align}
\end{description}

In the physical sector combination the Fock space ends up having positive definite
norm square and so this sector-combination is the one usual taken for being the in nature
realized one.

\subsection{Formulas for ``Majorana particles''}

The theory of Majorana fermions may be so well known that we had nothing to say,
but it were written about it in section 2.

For the boson case we introduced for each 
(vectorial) value of the momentum an operator acting
on the Fock space called $r(\vec{p})$ defined by (\ref{f281})
\begin{align}
r(\vec{p})&=\frac{1}{\sqrt{2}}\left(a(\vec{p},E>0)-a^{\dagger_f}(-\vec{p},E<0)\right)\nonumber
\\&=\frac{1}{\sqrt{2}}\left(a(\vec{p},E>0)+a^{\dagger}(-\vec{p},E<0)\right)
\end{align}
\\with the properties
\begin{align}
\left[r(\vec{p}),b(\vec{p'})\right]&=0\nonumber
\\\left[r(\vec{p}),b^{\dagger}(\vec{p'})\right]&=0\nonumber
\\\left[r(\vec{p}),b^{\dagger_f}(\vec{p'})\right]&=0
\end{align}
\\where the creation $b^{\dagger_f}(\vec{p})\left(=b^{\dagger_f}(\vec{p}\right)$ and
annihilation $b(\vec{p})$ operators for the ``Majorana bosons'' (i.e. boson being
its own antiparticle) were defined in terms of the $a$'s as 
\begin{equation}
b^{\dagger_f}(\vec{p})=\frac{1}{\sqrt{2}}\left(a^{\dagger_f}(\vec{p},E>0)+a(-\vec{p},E<0)\right)\label{f317}
\end{equation}
\\and
\begin{align}
b(\vec{p})&=\frac{1}{\sqrt{2}}\left(a(\vec{p},E>0)+a^{\dagger_f}(-\vec{p},E<0)\right)\label{f316}
\\&=\frac{1}{\sqrt{2}}\left(a(\vec{p},E>0)-a^{\dagger}(-\vec{p},E<0\right)
\end{align}

These operators obey (see(\ref{f280a}) and (\ref{f280b}))  
\begin{align}
\left[b(\vec{p},b^{\dagger_f}(\vec{p'})\right]&=\delta_{\vec{p}\vec{p'}}\nonumber
\\\left[b(\vec{p},b(\vec{p'}\right]&=0 \nonumber
\\\left[b^{\dagger_f}(\vec{p}),b^{\dagger_f}(\vec{p'})\right]&=0 
\end{align}
\\and so these operators are suitable for creating and annihilation of particles, and indeed
these particles are the ``Majorana bosons''. As a replacement for the in usual formalism for 
``Majorana bosons'' say
\begin{equation}
b(\vec{p},E)=b^{\dagger_f}(-\vec{p},-E)
\end{equation}
\\we have in our notation
\begin{equation}
b(-\vec{p})\mid_{\tiny\begin{matrix} with\\>\leftrightarrow <\end{matrix}}=b^{\dagger_f}(\vec{p})
\end{equation}
\\as is easily seen from (\ref{f316}) and (\ref{f317}) just above.

But now we need also a vacuum from which to start the creation of the 
``Majorana bosons'' with $b^{\dagger_f}(\vec{p})$.  In the two sector-combinations
1) the physical one and 4) ``a positive number with negative energy and vice versa''
there are the suitable vacua:

In 1)
\begin{equation}
\mid physical\ vac\rangle=\mid\ldots,k(\vec{p},E>0)=0,\ldots;\ldots,k(\vec{p},E<0)=-1,\ldots\rangle
\end{equation}

and in 4)
\begin{equation}
\mid\begin{matrix}pos\ in\ E<0\\neg\ in\ E>0\end{matrix}\rangle=1\ldots,k(\vec{p},E>0)=-1,\ldots;\ldots,
k(\vec{p},E<0)=0,\ldots\rangle
\end{equation}

In the sector-combinations 2) and 3), however, there are no charge conjugation symmetric states to use as the
vacuum state for a ``Majorana-boson'' formalism.  In this case the vacuum of 2) goes under charge 
conjugation $\bf{C}$ into that of 3).







\section{Outlook on String Field 
Theory Motivation}

One of our own motivations for 
developping the sort of boson Dirac sea 
theory for bosons being their own 
antiparticles,
i.e. a theory with Dirac sea, were to use 
it  in our own so called ``novel string field theory''\cite{1,2,Novel,
Nielsen:2014wna}. 

In this ``novel string field theory'' 
we sought to rewrite the whole of string 
theory\cite{12,13,14,15,21,18} (see 
also modified cubic theory \cite{9})-  although we did 
not yet come to superstrings\cite{10,17,11} although that should be relatively easy - 
into a 
formalism in which there seems a priori 
to be no strings.  The strings only come 
out of our
novel string field theory
\cite{3,4,5,6,7,8} by a rather 
complicated special way of looking at it.
  In fact
our basic model in this novel string field theory is rather like a system of /a Fock 
space for
massless scalar particles, which we 
call ``objects'' in our formulation, but 
they have much although not all 
properties similar to scalar massless 
particles.  These 
particles/objects  we think must be 
in an abstract way 
what we here called
Majorana bosons. This means they should 
be their own antiparticles to the extend 
that they have antiparticles. 

But their being put into cyclically 
ordered orientable chains may put a need 
for a 
deeper understanding of the Majorananess
for these ``objects''.

The reason for the objects, that in our 
novel string field theory are a kind 
of constituents, for the strings being 
supposed to a nature reminiscent of the 
Majorana particles or being their own 
antiparticles, is that they carry in 
themselves no particle number or charge, 
except that they can have (26)-momentum.
(For complete consistency of the bosonic
string theory it is wellknown that 26 
space time dimensions are required.)  
The bulk of the string (in string theory)
can namely be shrunk or expanded ad 
libitum, and 
it is therefore not in itself charged, 
although it can carry some conserved 
quantum numbers such as the momentum 
densities.

We take this to mean that the string as 
just bulk string should be considered to 
be equal to its own antimaterial. If we 
think of splitting up the string into 
small pieces like Thorn\cite{new12}, or we split the
right and left mover parts separately like
we did ourselves, one would in both cases 
say that 
the pieces of Thorn's or the objects 
of ours should be 
their own antiparticles.  With our 
a bit joking notation: they should be 
Majorana.  Thus we a priori could 
speculate that, if for some reason we 
should also like to think  
our objects as particles, then 
from the analytical properties of the 
single particle in relativistic theories
must have both positive and negative 
energy states. Then a treatment of 
particles being their own antiparticles 
in the Dirac sea formulation could 
- at least superficially -look to 
be relevant.   
 
One could then ask, what we learned above,
that could be of any help suggesting, 
how to treat long series of ``objects'',
if these objects are to be considered 
bosons that are their own antiparticles:

\begin{itemize}
\item{1.} In the novel string field theory of
ours it is important for the association 
to the strings, that one considers ring 
shaped chains of objects. We called such
ring shaped chains of objects for 
``cyclically ordered chains''. Now such 
ordering of our ``objects'' (as we call
them), or of any type of particles, into
chains in which each particle 
(or ``object'') can be assigned a number 
(although 
in our special model only a number modulo
some large number $N$) is o.k. for 
particles with an {\em individuality}. 
However, 
if we have particles (or ``objects'') that
are say {\em bosons}, then all particles 
are identical - or one could say any 
allowed state is a superposition of 
states in which all possible 
permutations on 
the particles have been performed and a 
superposition of the results of all these 
permutations with same amplitude only is presented as the final state  -. But this 
then means that  
{\em one cannot order them}, because you 
cannot say, which is before which in the 
ordering, because you cannot {\em name} 
the 
single particle. You could only say, that 
some particle A is, say, just before some 
particle B in a (cyclic) ordering, if you
characterize A as being the particle with 
a certain combination of coordinates (or 
other properties) and 
B as being the one with a certain other 
combination of coordinates (and other 
properties). Unless you 
somehow specify by e.g. some approximate 
coordinates (or other characteristic) 
which particle you think about, it 
has 
no 
meaning to express some relation involving 
the relative ordering, say, of two bosons. 

\item{2.} The problem just mentioned in 
assigning order to bosons means, that 
the concept of ``cyclically ordered 
chains'' of objects - or for that matter 
building 
up any string from particle pieces like
Thorn say - cannot be done once the 
particles or objects are bosons, but 
rather should
be preferably formulated {\em before one
symmetrize the wave function} under the 
particle permutation so as to implement 
that they are bosons. One shall so to
speak go back in the ``pedagogical'' 
development of boson-theory and think 
in the way before the symmetry principle 
under permutations making the particles 
bosons were imposed. In this earlier stage
of the description the cyclically ordered 
chains, or any type of ordering, which 
one might wish, makes sense. 
So here it looks that going back and 
postponing the boson constraint is needed
for ordering chains.

\item{3.} But seeking to go back prior to 
boson or fermion formulation makes a 
problem for the Dirac sea - in both 
boson and fermion cases -: If we want to 
consider the case of individual particles 
or objects fully, we have to imagine that 
we have given {\em names (or numbers) to 
all the 
particles 
in the Dirac sea}!   

For this problem we may think of 
a couple of solutions:
\begin{itemize}
\item{a.} We could imagine an interaction 
that would organize the particles in the 
ground state (to be considered a 
replacement for
the physical vacuum) or that some 
especially
important state for the Fock space 
obtained by imposing some  other 
principle is postulated to make up a kind 
of 
vacuum state. Then one could hope 
or arrange for the interaction or 
state-selecting principle chosen, 
that the vacuum state 
becomes such, 
that the objects (or particles)
in the Dirac sea goes into such a state, 
that these objects have such positions 
or momenta, that it due to this state 
becomes possible to recognize such 
structure 
that their ordering in the wanted chain
becomes obvious. If so, then 
the (cyclic) ordering 
can come to make 
sense.

This solution to the problem may be 
attractive a priori, because we then in 
principle using the now somewhat 
complicated state of the vacuum can assign
orderings to the whole Dirac sea, and thus 
in principle give an individuality even to 
the Dirac sea 
particles and missing particles 
/ the holes can make sense,  
too. They so to speak can inherit their 
individuality from the particles missing,
which before being removed were sitting 
 in the 
chains of the vacuum. We have thus at 
least got allowance to talk about a chain 
ordering for pieces of chains for the 
holes. There is so to speak an ordering 
of the holes given by the ordering of the
particles removed from the Dirac sea 
originating from the chain postulated to
have appeared from the interaction or from
some special selection principle for the 
vacuum state.

A little technical worry about the 
``gauge choice'' in our novel string field 
theory: In our novel string field theory
we had made a gauge choice for the 
parametrization of the strings, that led to
the objects having a special component 
of their momenta $p^+$, or in the language 
of our papers on this string field theory 
$J^+(I)$ for the $I$th object in the chain
{\em fixed} to a chosen small value
$a\alpha'/2$. Since the argument 
for there having to be negative energy
solutions(to say the Dirac equation) and
thus a need for a Dirac sea at all is 
actually analyticity of the equation of 
motion, we would suppose that also for
our objects one should keep ``analyticity''
in developing ones picture of the 
``negative energy states'' and thereby 
of the Dirac sea. But then the $p^+$ or
$J^+$, which is fixed to constant could 
hardly get continued to anything else 
than the same constant ? This sounds a 
bit unpleasant, if we imagine the $p^+$ 
be lightlike or timelike, because  then we 
cannot find the negative energy state with
the chosen gauge condition, and the whole
reason for the Dirac sea seems to have 
disappeared. And thus the discussion 
of Majorana may also have lost its ground.
But if we imagine the gauge choice fixed 
component to be spacelike, then we obtain, 
that the gauge condition surface 
intersects the light cone in two 
disconnected pieces that are actually 
having respectively positive and negative
energy. So assumming the gauge choice done
with a space-like component we have indeed
the possibility of the Majorananess 
discussion! And also in this case of
a spacelike $p$ or $J$ component being 
fixed (by gauge choice) our construction 
of the Majorana bosons makes perfect sense.

Now it gets again severely 
complicated by the chains postulated in 
the vaccuum.   In the space-like gauge 
fixing case it also becomes of course 
complicated, but the complication is 
due to the complicated state rather than 
to the gauge fixing alone.

Let us, however, stress again: To make 
a ordering of the objects in the Dirac 
sea into say cyclically ordered chains
a much more complicated state in the Fock 
space is needed than the simple say 
physical vacuum.

To figure out how to think about such a
situation with a ``complicated'' vacuum 
state replacing the, say, ``physical vacuum''
as discussed above, we might think about 
the analogous situation with the fermions.
When one has a quantum field theory with 
fermions having interactions, it means 
that the interaction part of the 
Hamiltonian has caused that the ground 
state for the full Hamiltonian is no 
longer the state with just the Dirac sea 
fillied and the positive energy single 
particle states empty. Rather it is 
a ``complicated'' superposition of states 
in the Fock space, most of which would 
in the free theory have positive energy.
These are states which can be described 
as states with some - infinite - number 
of positive energy fermions and some 
anti-fermions present (in addition to
the vacuum with just the Dirac sea 
filled). The presence of anti-fermions 
(holes) means, that if one acts with 
a creation operators $b^\dagger(\vec{p},E<0)$ 
for inserting a fermion with 
a negative energy $(E<0)$, then one 
shall not necessarily get $0$ as in the 
free theory vacuum, because one has the 
possibility(chance) of hitting a single 
particle state in which there is a {\em
hole}. The Fock-space state created 
by such an action will have 
higher full Hamiltonian energy than the 
``interaction vacuum'', because the 
latter is by definition the lowest 
energy state, but one has anyway 
succeeded in inserting a fermion 
in a state which from the free theory 
counted has a negative energy. 
It should be absolutely possible that 
such an inserted in the just mentioned 
sense negative energy particle could 
be part of the construction of say 
a bound state or some composite object 
resonance or so. Similarly it could on top 
of a ``complicated vacuum'' (meaning a 
ground state e.g. for the full 
Hamiltonian but not for the free one)
be possible to remove with an annihilation
operator $a(\vec{p}, E>0)$ a particle 
from a single particle state (having 
with the free Hamiltonian) positive energy
$(E>0)$. One could namely have the chance 
of hitting a positve energy single 
particle state, in which there already 
is a particle in the 
``complicated vacuum''. Such a removal 
or hole in a positive single particle 
state is what we ought to call 
a ``negative energy anti-particle''.     
 
We here sought to argue, that if one 
for some reason or another (because
of interaction and taking the ground state,
or because one has postuleted some 
``complicated vacuum'' just to make 
ordering make sense) use a ``complicated
vacuum'', then it becomes possible 
formally to add particles or 
anti-particles with {\em negative energy}.

Especially we want to stress the 
possibility
that, if one wants to describe properly 
a resonance or a bound state composed 
or several particles (e.g. fermions) 
then one might need to assign some 
of the constituents negative energy
in the sense just alluded to here.

Strictly speaking it comes to look 
in the ``complicated vacuum'' as if 
one has got {\em doubled the number 
of species of effective particle}, because
one now by acting with e.g. $a^\dagger(\vec
{p}, E>0)$ both can risk to produce 
a positive energy particle, and can risk
to fill in a hole in positive energy 
single particle state and thereby 
creating a
negative energy anti-particle. So 
operating with the same operator we risk
{\em two different results}, which may 
be interpreted as if one had effectively 
had two different types of operators and 
thereby doubly as many types of particles 
as we started with. We have so to
speak - in the case of non-Majorana
particles - gotten both positive and 
negative energy particles and 
also both positve and negative 
anti-particles effective on the 
``complicated vacuum''. 

If we go to make our particles Majorana,
we reduce the number of species by a 
factor two (as expected in as far as
Majorana means that particle and 
anti-particle gets identified.)

In the case of the ``complicated vacuum''
the transition to Majorana also
reduce the number of species by a factor 2
and thus compensates for the effect 
of the ``complicated vacuum''. With 
Majorana the particles and anti-particles 
are no longer distinguished, but with 
the ``complicated vacuum'' we obtain 
both positive and negative energy 
(Majorana)particles. It essentially functions as if the particle were no more 
Majorana. The ``complicated vacuum'', so to speak, removed the Majorananess.

We hope in later publication to be able 
to check that the just delivered story 
of the interaction vacuum increasing 
the number of species effectively by 
the factor two, is found when using 
the Bethe-Salpeter equation to describe
bound states. Then there ought according 
to the just said to be effectively both 
negative and positive states relevant 
for the ``constituent'' particles
in the Bethe-Salpeter equation.

Applying the just put forward point of 
view on the objects in our novel string 
field theory
we should imagine 
that in this formulation with the 
``complicated vacuum'' being one  with 
chains in it
it is possible for some objects to have 
their energy negative.
Nevertheless  
a whole 
chain formed from them might end up
with positive energy by necessity. 

Such a possibility of negative energy for
single objects that can nevertheless be
put onto the vacuum might be very 
important for complete annihilation of 
pieces of one chain put onto the vacuum 
with part on an other one also put onto
that vacuum. If we did not have such 
possibility for both signs along the 
chains, then we could not arrange that 
two incomming cyclically ordered chains
could partly annihilate, because energy
conservation locally along the chains 
would prevent that. 

At least in principle it must though be 
admitted, that such a picture based on an 
interaction  or by some restriction of the state of 
the whole world  makes a complicated 
vacuum is a bit  complicated technically.

But physically it is wellknown, that the 
vacuum in quantum field theories is 
a very complicated state, and so we might 
also expect that in string theory a 
similarly complicated vacuum would be 
needed. And that should even be the 
case in our novel string field theory 
in spite of the statement, often stated 
about this theory, that it has no 
interaction properly; all the seeming 
interactions being fake. But we could
circumvent the need for an interaction
to produce the complicated vacuum, we 
seemingly need by claiming that we instead
have a restriction on the {\em Fock space 
states}  of the system of objects, that
is allowed. Such a constraint could force 
the vacuum to be more complicated, and 
thus in succession lead to that it becomes allowed in the more complicated vacuum 
to have some of the objects having even 
negative energy, which in turn could 
allow a {\em complete annihilation }
of objects from one cyclicaally ordered 
chain and another set up in the same 
state (built on the complicated vacuum)

\item{b.} We give up seeing any chain 
structure in the vacuum as a whole, but
rather attempt to be satisfied with 
ordering the missing particles, (or may 
be the 
antiparticles?).

Naturally we would start imagining that 
we can have a Majorana boson, if we wish,
represented by -1 negative energy boson,
because the Majorana boson is a 
superposiotion of a particle and an 
antiparticle,
and the latter really can be considered 
-1 particle of negative energy.

At first one might think that having 
two bound states or two strings, which 
would like to partially annihilate -as
it seems that we need in our derivation 
of Venezianoamplitude in our novel 
string field theory - could be indeed 
achieved by having part of one of these 
composed structures treated or thought 
upon as consisting of antiparticles, 
since one would say that particle and anti-particle can annihilate. However, 
when anti-particle and particle both 
with positive energy annihilate, then 
at least some energy is in excess and 
they therefore cannot annihilate 
completely into nothing. Rather there
would have to some emmitted material 
left over to take away the energy.
If we therefore as it seems that we 
would to get the terms missing in 
our novel string field theory t get 
the correct three term Veneziano 
amplitude should have a total annihilaton without left over such positive energy 
particles and antiparticles are not
sufficient. Therefore this b. alternative
seems not to truly help us with the 
problem of our novel string field theory
to reproduce the Veneziano model fully.    
 \end{itemize}

\item{4.} In our formalism above - taken 
in the 
physical vacuum -  the 
``Majorana-boson'' became a superposition 
 of being a hole and a genuine positive 
enrgy particle. The hole meant it were 
in part of the superposition - i.e. 
with some probability 50 \% - $-1$ 
particle with negative energy.  So one 
would with significant probability be 
able
to consider that the ``Majoran-boson''
were indeed a lack of a negative energy
original particle. For calculating 
amplitudes of some sort one would then 
imagine that we might even have to add 
up contributions from the holes and 
contributions from the positive energy 
particles.

For each object, say, we should think we 
should have both a contribution in 
which it is considered a particle (with 
positive energy) and one in which it 
is a hole. 
\item{5.} From the construction of the 
creation and annihilation operators for 
the ``Bosons being their own 
antiparticles
'' - the $b$'s - being constructed as 
containing the quite analogous 
contributions from a hole part and 
a particle part, it 
looks that in building up states with 
many 
Majorana- bosons one gets an analogous 
built up for both the holes and the 
particles and with say the analogous 
momenta.
 
Here analogous means that the holes are 
holes for states with opposite momentum, 
but since it is holes it becomes the 
same 
net momentum for the hole as from the 
particle analogous to it.

\item{6.} With any sort of even formal 
interaction one would think that a hole
and a particle can annihilate as stuff
annihilate anti-matter. But if you have 
a pair of positive energy particles or 
anti-particles, they can only
annihilate into some other particles of 
some sort. They cannot just disappear
together. That is however, possible, 
if you have a negative energy particle and
a positive energy one of just opposite 
four(or 26) momenta. 

\item{7.} If one would say choose a 
gauge so that the particles get as in 
our gauge choice in our papers on the 
novel string theory that a certain 
momentum  
component, $p^+$ say, is specified to be 
a fixed value $a\alpha'/2$ as we choose,
then one would have to let the 
particle, the state of which is made the 
hole have its $p^+ = -a\alpha'/2$, i.e. 
the opposite value. (Then if 
one has negative numbers of such particles, of course they contribute a positive 
$p^+$ again.) If one has indeed completely
opposite four momenta 
- including energy - 
then an anihilation without left over 
is possible, otherwise not. It is 
therefore it is so crucial with negative 
energy constituents, if any such total 
disappearance of a pair is needed/wanted. 

But if we have physically only the 
free simple vacuum in which one has just 
for bosons emptied the negative energy 
states and for fermions just filled the 
negative energy states and no more, then 
all modifications will even particle for 
particle have positive energy. It will
either be a removal of a negative energy
particle meaning an antiparticle created
or an insertion of a positive energy 
particle. Both these modifications would 
mean insertion of  positive energy and 
they could not annihilate with each other without leaving 
decay material. So to have a piece of a 
cyclically ordered chain annihilate 
{\em without
decay material} with another piece, it is 
needed that we {\em do not just have 
the free  theory vacuum}. We need instead
something like a ``complicated vacuum''
such as can be gotten by the effect 
of either interactions, or from some more
complicated postulate as to what the vacuum state should be. 

In our novel string field theory, in which it is claimed 
that there are no interactions in the object formulation, we cannot 
refer to interactions. Rather we 
must refer to making a postulate about 
what the ``complicated vacuum state'' 
should 
be. As already mentioned above we need 
in order that ordering into the 
cyclically ordered chains can make 
sense to have 
as the (vacuum) state a state in which 
the various objects can have so different
single particle states that we can use 
their single particle state 
characteristic to mark them so as to 
give them 
sufficient individuality. Really we should
postulate such a ``complicated vacuum 
state'' that there would for each object
be an effectively unique successor lying 
as neighbor for the first one. But such 
restrictions to somewhat welldefined 
positions relative to neighbors in a chain
must mean that it cannot at all be so, that
there are just, 1 for fermions, -1 for 
bosons, particels in the negative energy 
states and  zero in the positive ones.
Rather it means, that considering such 
a free vacuum as starting point the 
state with the chains organized into the 
``compicated vacuum'' is strongly excited.
So there are many both particles and anti
particles present in this ``complicated
vacuum '' needed to have chains inside 
the 
vacuum.

But as already said such ``complicated
vacuum'' can give the possibility of 
having effectively negative energy 
constituents. Since our objects are 
essentially
constituents, this also means that our 
objects in a complicated vacuum can 
get allowed to be of negative energy.
We must arrange that by allowing them 
in our gauge choice to get the 
$J^+$ have both signs. If so we may 
enjoy the full annihilation without 
left over material.

\item{8.} To construct an operator 
creating 
a chain (or series) of Majorana particles 
- in our novel SFT we mean the objects - 
we strictly speaking should use a 
specific
linear combination of the hole and the 
positive energy particle (or object) for
every Majorana particle created along the 
chain, but if we project out at the end 
the constructed Fock space state into 
the subspace used for the Majorana 
boson description, it is not so important
to use precisely the correct linear 
combination. We shall namely obtain the 
 right linear combination, since
in that case it comes  out of such 
a projection automatically. 

But trusting that projecting into the 
Majorana-describing sub-space will do 
the job, we can just choose at will 
whether
we use a series of positive energy 
particle (or object) creation operator
or instead the corresponding hole creating
(destruction of negative energy) operator. 
\end{itemize} 
   

Since our objects are a priori Majorana
ones, it  may at the end due to the 
doubling of state-types mentioned get 
them rather 
described effective as non-Majorana, in 
the way that they can be in both 
positive and negative energy single 
particle states. 
This actually reminds us  more about the 
``naive vacuum'' sector combination.
But now it is the result of the 
``complicated vacuum'' and of the thereby
associated ``doubling of the number 
of species effectively''.

\subsection{The ``Rough Dirac Sea'' 
in General} 
Let us extract and stress the idea, which 
we suppose will be very important for 
our formulation of the scattering 
amplitude for strings in our novel string
field theory, but which could also be 
imagined to deliver an approximation that 
could be useful especially for bound 
states with many constituents, ``the 
(very) rough Dirac sea''. This rough 
Dirac sea is really the same as what we
called above the ``complicated vacuum''.

The picture of true rough sea 
(a rough sea is the opposite of a calm
sea, and it means that there lots of
high waves
may actually)
be a very good one to pedagogically 
promote the idea of the effects of the 
``complicated vacuum'' or the ``rough 
Dirac sea'' leading to that we effectively
get negative energy particles and antiparticles.

In this picture the ``calm Dirac sea''
means the free approximation vacuum, in
which - in the physical choice of sector
combination, which is what one  
normally will have in mind - the negative
energy states are filled for the fermion 
case, while ``emptied out'' in the boson 
case. In any case this calm Dirac sea
is the picture for the theory vacuum in
the unperturbed approximation 
(the free vacuum). But in interacting 
quantum field theories the vacuum gets 
perturbed by the interaction and becomes 
a more complicated state ``the complicated
vacuum'', and it is for this ``complicated
vacuum'' that the analogy with the 
rough sea is very good. There  
should have been near the surface -at  the average 
surface height - a region in heights, in 
which you find  with some probability
water and with some probability air. Just 
at  
the should-have-been surface (= average
surface) one expects that the probability 
for finding water is 50 \% and for finding
air in a given point is 50 \%. 

Now imagine: we come with an extra 
water molecule (or may be just a tiny bit
of water) and want to insert it into 
the sea or the air not too far from the 
``should-have-been surface''. Now if 
there happen
to be a wave of water present, where you 
want to or attempt to insert such an 
extra tiny bit of water, you will not 
succeed, and
that is analogous to getting zero, when 
you want to create a particle with a 
creation 
operator into a state that is already 
filled (say, we think for simplicity on 
the fermion case). If, however, there 
happen
to be a valley in the waves, you will 
succeed in inserting a tiny bit of water
{\em even if it is under the average 
water height}! This corresponds to 
inserting a negative energy particle into 
the 
``rough Dirac sea'' or the ``complicated
vacuum''. You may also think about 
removing a droplet of water. That will of 
course only succeed, if there {\em is} 
some
water in the point in space, wherein you 
want to do it. Again it is not guaranteed 
that you can remove a bit of water in the 
rough sea, even if you attempt to remove 
it deeper than the average water height,
because there might be a valley among the 
waves.
Also if you hit a wave you might be able 
to remove a bit of water from a height 
above the average height.

In this way we see that you can produce 
sometimes a hole in the water both with
positive and negative height (analogous to
the both positive and negative (single particle) energy). Similarly you may produce 
both above and below extra bubbles of water.

This means that we have got a kind of 
doubling: While in the calm Dirac sea 
you can only make droplets (~ particles)
above the average surface and only holes
(~ antiparticles) below, we now in {\em 
the 
rough sea can do all four combinations}.
    
\subsection{Infinite Momentum Frame Wrong,
in Rough Dirac Sea?}

With ``rough Dirac sea''-thinking we
arrived at the idea, that one might 
describe for instance a bound state 
or resonance as composed of constituent 
particles not all having positive 
energy; but some of the constituents 
could have {\em negative} energy.

It must be legal to choose to describe 
a bound state or resonance state by 
a linear combination - weighted with 
what is essentially a wave function for 
the constituents in the bound state or 
resonance - of creation operators and 
annihilation operators (for describing 
the contained anti-particles among the 
constituent particles) and let it act 
on the vacuum. We might, say, think 
of an operator of the form
\begin{align}
&A^\dagger(\hbox{bound state})=\\
&= \int 
\Psi(\vec{p}_1, h_1,s_1; ...; 
\vec{p}_N,h_N,s_N)\nonumber \\
&*\prod_{h_1,s_1} (a^\dagger(\vec{p}_1,
h_1,s_1) d^3\vec{p}_{1,h_1,s_1})\cdots 
\prod_{h_N,s_N} 
(a^\dagger(\vec{p}_n,h_N,s_N)d^3\vec{p}_N);\\
&|\hbox{bound state (Fock)state}\rangle=\\
&= A^\dagger(\hbox{bound state})|\hbox{``complicated vacuum''}\rangle\\
&= \int 
\Psi(\vec{p}_1, h_1,s_1; ...; 
\vec{p}_N,h_N,s_N)\\
&\prod_{h_1,s_1} (a^\dagger(\vec{p}_1,
h_1,s_1) d^3\vec{p}_{1,h_1,s_1})\cdots 
\prod_{h_N,s_N} 
(a^\dagger(\vec{p}_n,h_N,s_N)d^3\vec{p}_N)\\
&|\hbox{``complicated vacuum''}\rangle
\label{wffbs}
\end{align}
where $\Psi(\vec{p}_1, h_1,s_1; ...;\vec{p}_N,h_N,s_N)$ 
is (essentially) the
wave function for a bounds state of $N$
constituents numbered from $1$ to $N$.
The momenta of the constituents are 
denoted by $\vec{p}_i$ with $i=1,2,...,N$,
while the internal quantum numbers are 
denoted $h_i$, and then {\em there is 
the symbol $s_i$ that can be $s_i =$ ``positive''$= (E>0)$ or $s_i=$``negative'' 
$= (E<0)$ }, meaning that the single 
particle energy of the constituent here is 
allowed to be both positive and negative,
it being denoted by $s_i$, which of these 
two possibilities is realized for constituent number $i$. In this expression 
(\ref{wffbs}) we took just $N$ 
constituents, but it is trivial to write 
formally also the possibillity of the 
bound state being in a state, that is 
a superposition of states with different 
values of the number $N$ of constituents: 
\begin{align}
&A^\dagger(\hbox{bound state})=\\
&= \sum_{N=1,2,...}\int \Psi_N(\vec{p}_1, h_1,s_1; ...; 
\vec{p}_N,h_N,s_N)\nonumber \\
&*\prod_{h_1,s_1} (a^\dagger(\vec{p}_1,
h_1,s_1) d^3\vec{p}_{1,h_1,s_1})\cdots 
\prod_{h_N,s_N} 
(a^\dagger(\vec{p}_n,h_N,s_N)d^3\vec{p}_N).
\end{align}
In this way we could describe a (bound)
state inserted on the background of the 
true (``complicated'') vacuum with 
a superposition of different numbers 
of constituents. In principle we could 
find a wave function set, $\Psi_N(\vec{p}_1, h_1,s_1; ...;\\ 
\vec{p}_N,h_N,s_N)$ for $N=1,2,...$, that
could precisely produce the (bound) state
 or resonace in question. It might because 
of the allowance of both negative and 
positive energy constituents be possible 
to construct in this way a given state 
in more than one way. But one could well
imagine, that if we would like to have the
wave function reasonably smooth, then 
it would be hard to quite avoid the 
negative energy constituents contributions - they are 
of course only relevant by giving nonzero
contributions to the state created 
provided the Diarc sea is rough - and 
thus it looks like being essentially 
needed to use wave functions with also 
negative energy constituents, unless  
one is willing to give up the 
accuracy in which the influence from 
the interaction on the vacuum must be 
included.

But if we thus accept a description with
negative constituent energy, the usual
thinking on the 
``infinite momentum frame''\cite{IMF} seems wrong:

If we in  fact have constituents with 
single particle state negative energy,
then boosting such a state eversomuch 
 in the longitudinal momentum
direction cannot bring these negative 
energy constituents to get posive 
longitudinal and thereby positive 
Bjorken $x$. So the usual story that 
provided we boost enough all constituents
obtain positive $x$  cannot be kept in
our rough Dirac sea scenario with its
negative energy constituents!

This may be the reason for the trouble in 
our novel string field theory which 
triggered us into the present work. In 
this 
novel string field theory formulation 
we namely used infinite momentum frame 
and actually took it, that all the 
there called objects -  which are 
essentially constituents -  had their
$J^+= a\alpha'/2$. But now the 26-momentum,
which is proportional to the $J^{\mu}$, 
should then for all the objects have the 
$+$ component positve. But now the 
notation is so, that this $+$ component 
means the longitudinal momentum in the 
infinite momentum frame. So we assumed 
a gauge choice in our formulation 
of this novel string field theory which
is inconsistent with the negative energy 
constituent story arising from rough 
Dirac sea.

This ``mistake'' is very likely to be the
explanation for the strange fact, that 
we in deriving the Veneziano model
from our novel string field theory 
formalism  
only got {\em one out of the three terms 
we 
would have expected.}

The suggested solution to our trouble 
would then be to allow also for 
constituents with the $J^+$ being negative.
That would mean we could not keep to 
the simple gauge choice enforcing 
a positive value to $J^+$ but would have
to allow also negative values for this 
$J^+$.

That in turn might then allow constituent 
pairs from say different bound states
- or different strings as it would be in 
our formalism - to totally annihilate 
meaning without leaving any material 
after them, because no excess energy 
would have to be there after the 
annihilation. Negative energy and 
positive energy together have the chance 
of such total annihilation.




\section{Conclusion and Outlook}


The in many ways intuitively nice and appealing language of the Dirac sea, which we
have in an earlier work extended also to be applicable for bosons, is at first
not so well suited for particles --``Majorana particles''-- which are identical
to their own antiparticles.  In the present article we have nevertheless
developped precisely this question of how to describe particles --bosons or fermions--
which are, as we call it, ``Majorana''. We use also this terminology ``Majorana''
even for bosons to mean that a particle is its own antiparticle.  The fermion case
is rather well known.  So our main story was first to review, how it
were at all possible to make (a free) theory for bosons based on a Dirac sea, and
secondly the new features of this Dirac sea for boson theory as follows:
\begin{description}
\item{a)}negative norm squares
\item{b)}negative number of particles.
\end{description}

The main point then became how to get what we call a Majorana-boson theory
through these new features.  This comes about by constructing in terms of the
creation and annihilation operators $a^{\dagger_f}(\vec{p},E>0)$ and $a(\vec{p},E<0)$
for a type of boson that might have a charge, some creation and annihilation
operators $b(\vec{p})$ and $b^{\dagger_f}(\vec{p})$ for the Majorana boson, which
is really a superposition of a boson and an anti-boson of the type described by
$a$ and $a^{\dagger}$.

\subsection{The old Dirac sea for bosons}

The Dirac sea for boson theory is based on having a Fock space, for which a basis
consists of states with a number of bosons $k(\vec{p},E\gtrless0)$, which can be
both positive, zero and \underline{negative} integer, in both positive and
negative energy $E$ single particle states for each $3-momentum\ \vec{p}$,
\begin{equation}
\mid\ldots,k(\vec{p},E>0),\ldots;\ldots,k(\vec{p'},E<0),\ldots\rangle\in Fock\ space.
\end{equation}

Because of the complication that the inner product 
\begin{equation}
\langle\varphi_1\mid_{(f)}\varphi_2\rangle=\int \varphi^\ast_1\frac{\overleftrightarrow{\partial}}{\partial_t}\varphi_{2}d^{3}\vec{X}
\end{equation}
\\for a (single particle) boson is \underline{not} positive definite we have to distinguish two different inner
products $\mid$ and 
$\mid_f$ say and thus also the two thereto responding hermitean conjugations
$\dagger$ and $\dagger_f$, meaning respectively without and with the 
$\int \varphi^\ast_1\frac{\overleftrightarrow{\partial}}{\partial_t}\varphi_{2}d^{3}\vec{X}$ included.
In fact we have for the norm square for these two inner products
\begin{align}
&\langle\ldots,k(\vec{p},E>0),\ldots;\ldots,k(\vec{p'},E<0),\ldots\mid_f\ldots,k(\vec{p},E>0),\ldots;\ldots,k(\vec{p'},E<0),\ldots\rangle\nonumber
\\=&(-1)^{\sharp(neg.\ energy\ b.}\langle\ldots,k(\vec{p},E>0),\ldots;\ldots,k(\vec{p'},E<0),\ldots\mid\ldots,k(\vec{p},E>0),\ldots;\nonumber
\\&\ldots,k(\vec{p'},E<0),\ldots\rangle\nonumber
\\=&(-1)^{\sharp(neg.\ energy\ b.}\cdot\prod_{(\vec{p},E\gtrless 0)for\ which\ k\le-1}(-1)^{|k|}
\end{align}


\subsection{Main Success of Our Previous Dirac Sea (also) for Bosons:}
The remarkable feature of the sector with 
the emptied out Dirac sea for bosons 
- what we called the physical sector -
is
that one has  arranged the sign 
alternation  (\ref{alt})
with the total number of negative energy 
bosons to cancel
 the sign from (\ref{inb}) so as to
achieve 
 that the  
total Fock space has positive norm 
square.  This ``physical sector'' 
corresponds 
to that negative energy single particle 
states
are in the negative sectors, while the 
positive 
energy single particle states are in the 
positive
sector.

Thus the basis vectors of the full Fock 
space for the physical
sector are  of the form
\begin{equation}
|\ldots,k(\vec{p},E>0),\ldots;\ldots,k(\vec{p},E<0),\ldots \rangle\label{basis}
\end{equation}
\\where the dots \ldots denotes that we have one integer
number for every momentum vector 
--value ($\vec{p}$ or $\vec{p'}$),
but now the numbers $k(\vec{p},E>0)$ of 
particles in a positive
energy are-- in the physical sector-combination-- restricted to be
non-negative while the numbers of bosons in the negative energy
single particle states are restricted to be negative
\begin{eqnarray}
k(\vec{p}, E>0)=&0,1,2,\ldots \nonumber
\\k(\vec{p}, E<0)=&-1,-2,-3,\ldots
\end{eqnarray}

{\bf In this physical sector our Dirac Sea
formalism is completely equivalent to the 
conventional formalism for quantizing 
Bosons with ``charge'' (i.e. Bosons that 
are not their own antiparticles), say e.g. $\pi^+$ and 
$\pi^-$.}

But let us remind ourselves that this 
idea of using Dirac sea {\bf allows one to 
 not fill the Dirac sea}, if one should 
wish to think of such world. With our 
extension of the idea of the Dirac sea 
to also include Bosons one also gets 
allowed to not empty out to have $-1$
boson in each negative energy single 
particle state. But for bosons you have 
the further strange feature of the 
phantasy world with the Dirac sea not treated as it should be to get physical, that 
one even gets negative norm square states,
in addition to like in the fermion case 
having lost the bottom in the energy.

\subsection{Present Article Main Point were to Allow for Bosons being their own Antiparticles also in Dirac sea Formalism}

We could 
 construct a ``Majorana-boson'' creation operator for say
a ``Majorana-boson'' with momentum $\vec{p}$, $b^{\dagger}(\vec{p})$
analogously to the expressions (\ref{cref}) and (\ref{anf}).
\\$b^{\dagger}(\vec{p})=\frac{1}{\sqrt{2}}\left(a^{\dagger}(\vec{p},E>0)+a(-\vec{p},E<0)\right)$
and $b(\vec{p})=\frac{1}{\sqrt{2}}\left(a(E>0)+a^{\dagger}(E<0)\right)$

Since an extra phase on the basis states does not matter so much we could
also choose for the bosons the ``Majorana boson'' creation and annihilation
operators to be 
\begin{align}
b^{\dagger_f}(\vec{p})&=\frac{1}{\sqrt{2}}\left(a^{\dagger}(E>0)+a(-\vec{p},E<0)\right)\nonumber\label{f228c}
\\&=\frac{1}{\sqrt{2}}\left(a^{\dagger_f}(\vec{p},E>0)+a(-\vec{p},E<0)\right)\nonumber
\\b(\vec{p})&=\frac{1}{\sqrt{2}}\left(a(\vec{p},E>0)+a^{\dagger_f}(-\vec{p},E<0)\right)\nonumber
\\&=\frac{1}{\sqrt{2}}\left(a(\vec{p},E>0)-a^{\dagger}(-\vec{p},E<0)\right)
\end{align}

Such creation  operators $b^{\dagger_f}(\vec{p})$ and their corresponding annihilation 
operators $b^(\vec{p})$ make up the 
completely usual creation and annihilation 
operator algebra for Bosons that are 
their own antiparticles in the case of the
``physical sector combination''. This 
``physical sector combination'' means 
that we emptied out the Dirac sea in the 
sense that in the ``vacuum'' put just $-1$ boson in each negative energy 
single particle state. This correspondence
means that our formalism is for this 
``physical sector combination'' completely
equivalent to how one usually describes 
Bosons - naturally without charge - which 
are their own antiparticles. But our 
formalism is to put into the framework 
of starting with a priori ``charged''
Bosons which then quite analogously to 
fermions have the possibility of having 
negative energy (as single particles). 
We then treat the analogous problem(s) 
to the Dirac sea for Fermions, by ``putting {\em minus} one boson in each of the 
negative energy single particle states.
That a bit miraculously solves both 
the problem of negative norm squares and negative second quantized energy, and even
we can on top of that {\em restrict 
the theory}, if we so wish,{\em to 
enforce the bosons to be identified 
with their own antiparticles.}

We saw above that 
\begin{itemize}
\item{1.} We obtain the Fock-space 
(Hilbert-space) for the Bosons being 
their own antiparticles by 
{\em restriction} to a subspace
\begin{equation}
H_{Maj}=\left\{\mid \rangle \mid r(\vec{p}) \mid \rangle=0\right\},
\end{equation}
where we have defined
\begin{eqnarray}
r(\vec{p})&=&\frac{1}{\sqrt{2}}\left(a(+\vec{p},E>0)+a{\dagger}(-\vec{p},E<0)\right)\nonumber
\\&=&\frac{1}{\sqrt{2}}(a(\vec{p},E>0)-a^{{\dagger}_f}(-\vec{p},E<0).
\end{eqnarray}
Of course when one forces in the 
original Dirac sea formalism  
the antiparticles differnt from the particles 
to behave the same way in detail it means
a drastic reduction of the degrees of 
freedom for the second quantized system
- the Fock space-, and thus it is 
of course quite natural that we only use 
the subspace $H_{Maj}$ being of much less
(but still infinite) dimension than the 
original one. 

\item{2.} In our formalism - since 
we use to write the creation operator 
for the boson being its own antiparticle 
as a sum of creation of a particle and of 
a hole (\ref{f228c}) - a ``Majorana-boson''is physically described as statistically 
or in superposition being with some chanse
a particle and with some chance a hole.
Really it is obvious, that it is $50$ \%
chance for each. So the physical 
picture is that the ``Majorana-boson''
is a superposition of a hole and an
original positive energy particle in the 
``physical sector combination''.
\item{3.} We could construct a charge 
conjugation operation ${\bf C}$ which 
on our Fock space with both negative and 
positive energy states present as possibilities obtained the definition:
\begin{align}
{\bf C}\mid& \ldots \stackrel{\sim}{k}(\vec{p},E>0),\ldots;\ldots,\stackrel{\sim}{k}(\vec{p'},E<0),\ldots\rangle\nonumber\label{cdef}
\\=\mid&\ldots,k(\vec{p},E>0)=-\stackrel{\sim}{k}(-\vec{p'},E<0)+1,\ldots;
\\
&\ldots,k(\vec{p'},E<0)\nonumber
=-1-\stackrel{\sim}{k}(-\vec{p'},E>0),\ldots\rangle.
\end{align}
Of course the state of the system of 
negative single particle comes to depend 
on that of the positive energy system 
after the charge conjugation and 
oppositely. With (\ref{f228}) or (\ref{f228c}) one sees that on the whole system 
or Fock space of Bosons being their  own
antiparticles is left invariant under 
the charge conjugation operator ${\bf C}$.
This is as expected since these ``Majorana
Bosons'' should be invariant under 
${\bf C}$.  
\end{itemize}
\subsection{The Unphysical Sector Combinations and Boson-theories therein with Bosons being their own antiparticles}
As a curiosity - but perhaps the most new 
in the present article - we have not only 
the physical sector combination, which so
successfully just gives the usual 
formalism for both ``charged''  bosons and
for what we called Majorana-bososns (the 
ones of their own antiparticles) but three 
more ``sector-combinations'' meaning combinations of whether one allows only 
negative numbers of bosons, or 
only non-negative numbers for the positive 
and the negative single particle states.
The reader should have in mind that there 
is what we called the barier, meaning that
the creation and annihilation operators 
cannot cross from a negative number of particles in a single particle state to a positive one or opposite, and thus we can 
consider the theories in which a given 
single particle state has a positive or zero number of particles in it as a 
completely different theory from one in 
which one has a negative number of bosons 
in that single particle state. For simplicity we had chosen to only impose that 
we only considered that all single 
particle states with one sign of the 
single particle energy would 
have  their  number of particles being 
on the same side of the barrier. But even 
with this simplifying choice there remained
$2^2=4$ different sector-combinations.
One of these sector-combinations - and 
of course the most important one because
it matches the usual and physical 
formalism - were the ``physical 
sector combination'' characterized by
their being a non-negative number $k$ of 
bosons in all the positive energy single particle states (i.e. for $E>0$), while the
number $k$ of bosons in the negative 
energy 
single particle states (i.e. for $E<0$)
is restricted to be genuinely negative 
$-1,-2,-3,...$.

The sector combination possibility 4) in our enumeration 
above the Fock space gets negative 
definite instead of as the one for the 
physical sector combination which gets 
positive definite. But these sector combinations are analogous or isomorphic with the appropriate sign changes allowed. Also our 
charge conjugation operator ${\bf C}$ 
operates inside both the ``physical
sector combination'' and inside the 
sector combination number 4), which is 
characterized as having just the opposite 
to those of the physical sector, meaning 
that in sector combination 4) one has a 
negative number of particles in each 
positive energy (single particle)state,
while there is a positive or zero number 
in the negative energy states. Thus the 
construction of particles being their 
own antiparticles would be rather 
analogous to that in the physical sector 
combination. 

Less trivial is it to think about the two
sector combinations 2) and 3) because 
now the charge conjugation operator 
${\bf C}$ {\em goes between them}:Acting 
with the charge conjugation operator 
${\bf C}$ on a state in the section combination 2) which we called the ``naive 
vacuum sector combination'' one gets 
a result of the operation in the 
{\em different sector combination} namely 
3). You can say that the charge 
conjugation operator does not respect 
the barrier, it is only the creation 
and annihilation operators which respect 
this barrier. A priori one would therefore now expect that one should construct the 
formalism for the boson being its own antiparticle for these sector combinations 2) and 3) based on a Fock space covering both 
parts of the sector combination 2) and 
part of 3). To realize that one gets 
eigenstates of the charge conjugation 
operator such a combination of the the two 
sector combinations is of course also 
needed. However, if one just wanted to 
realize an algebra of the creation and annihilation operators that could be 
interpreted as a formalism for the boson
type being its own antiparticle, one might throw away one of the two sector 
combinations, say combination 3), and 
keep only the ``naive vacuum sector 
combination'' 2). Since the creation and 
annihilation operators can{\em not} cross
the barrier from one sector combination into the other one, such a keeping to only 
one of the two sectors between the charge conjugation operator goes back and forth
would not make much difference for the 
creation and annihilation operators. 
We did in fact develop such a formalism 
for bosons being their own antiparticles 
in this way in alone ``the naive vacuum 
sector combination''. Interestingly it now
turned out that keeping to only one 
sector combination the whole Fock space 
constructed for the boson being its own antiparticle became of zero norm square.
Really we should say Hilbert inner product 
became completely zero for the subsector 
of the Fock space - of this unphysical 
``naive vacuum sector combination'' -.
This is of course at least possible since 
the sector combinations 2)(=the naive vacuum one) and 3) have both positive and 
negative norm square states- so that 
no-zero Hilbert vectors can be formed as
linear combinations of positive and negative normsquare Hilbert-vectors. (In the 
physical sector combination nor the 
sector combination 4) zero norm states 
cannot be found because the Hilbert 
innerproduct is respectively positively
and negatively definite.).          
     
\subsection{Speculations Bound States, Rough Dirac Sea etc.}

Then in the last section above we have 
some to the rest more weakly connected 
speculations meant to be of help for the 
original problem bringing us to the considerations in this article, namely our ``novel 
string field theory''. A major suggestion, that came out of these considerations were 
to have in mind that, when you have an 
interacting quantum field theory, the 
vacuum gets into a rather complicated 
superposition of Fock space states, that
makes  descriptions as the 
``rough Dirac sea'' or ``the complicated
vacuum'' appropriate. While in say 
the ``physical vacuum'' - descussed in the 
article - you can only remove particles 
from negative energy states and only add 
particles to the positive single particle 
states, one does {\em not} have this 
restriction in the interacting vacuum,
or say the ``rough Dirac sea'' vacuum.
This point of view suggests that to make 
a proper description of a bound state or 
a resonance by means of a wave function
in a relativistic quantum field theory,
describing how to add or remove 
constituents from the ``rough Dirac sea''-vacuum one should include {\em also 
negative energy possibilities} for the 
particles or antiparticle constituents. 

These considerations are also hoped to be 
helpful for the problems we have for the 
moment with obtaining the full Veneziano
model amplitude from our novel string 
field theory.

  

\section*{Acknowledgement}
One of us (H.B.N.) wishes to thank the 
Niels Bohr Institute to allow him to work 
as emeritus there, with room and some support for travel etc. The other (M.N.)
acknowledges the Niels Bohr Institute and the Niels Bohr International Academy,
where the present work started;  He wishes to thank there for their hospitality 
extended to hime during his stay.  M.N. also acknowledges the Okayama Institute
for Quantum Physics and the Okayama local government for providing him travel expences.
He also acknowledges that present research is supported by the J.S.P.S. Grant-in-Aid for 
Scientific Research No. 23540332.

\end{document}